\documentclass[aps, pra, a4paper, showpacs, twocolumn, english, 10pt, nofootinbib]{revtex4-2}

\usepackage{graphicx}
\usepackage{dcolumn}
\usepackage{bm}

\usepackage{braket}
\usepackage{xspace}
\usepackage{tikz,quantikz}
\usepackage{orcidlink}
\usepackage{comment}
\usetikzlibrary{external}
\usepackage{amssymb}
\usepackage[T1]{fontenc}
\usepackage{subfigure}
\usepackage[normalem]{ulem}

\usepackage[capitalise]{cleveref}	

\usepackage{import}
\usepackage{transparent}

\usepackage[capitalise]{cleveref}	

\usepackage[mode=build]{standalone}
\newcommand{\ea}{\textit{et al.}\xspace}

\usepackage{graphicx}
\usepackage{dcolumn}
\usepackage{bm}

\crefname{appendix}{Appendix}{Appendices}
\crefname{section}{Section}{Sections}

\begin{document}

\preprint{APS/123-QED}
\title{Limitations of Error Model Approximations in Quantum Network Simulation}
\author{Julia Freund\orcidlink{0000-0001-5548-5007}$^{1}$, Jorge Miguel-Ramiro\orcidlink{0000-0001-9723-7298}$^{1}$, Julius Wallnöfer\orcidlink{0000-0002-4837-2757}$^{1}$ and Wolfgang Dür\orcidlink{0000-0002-0234-7425}$^1$}
\affiliation{$^1$ Universität Innsbruck, Institut für Theoretische Physik, Technikerstraße 21a, 6020 Innsbruck, Austria}

\date{\today}

\begin{abstract}
Efficient classical simulation of large-scale quantum networks frequently relies on noise approximations, which consider a restricted set of operators to describe noisy channels and operations. In this work, we demonstrate how such simplified error models, such as Pauli twirling or reset channels, can lead to severe quantitative and qualitative discrepancies in protocol performance predictions. We analyze, in particular, how small differences can accumulate in iterative and sequential protocols such as entanglement purification, entanglement swapping, and repeater chains. Our results reveal that neglected error contributions can lead to important performance under- and over-estimations, measurement-outcome dependency, and oscillations in the fidelity, which are entirely overlooked by the simplified error model approximations. These results show that rigorous validation of complete noise architectures is indispensable for accurately predicting operational thresholds in future quantum technologies.
\end{abstract}
\maketitle


\section{Introduction}
Quantum networks promise to be the backbone of future quantum technologies~\cite{NOISE:Internet_0,Kimble2008, Wehner2018}, enabling not only secure communication~\cite{NOISE:ShorQKD,NOISE:GottesmanQKD,NOISE:ZhaoQKD,NOISE:Murta_2020_quantum_conference,NOISE:ACKA}, but also unique applications such as distributed quantum computing~\cite{CiracDistributed, Hayashi15, Cacciapuoti2020}, blind quantum computation~\cite{Barz2012, Fitzsimons2017}, or distributed sensing~\cite{Giovannetti_2011, Sekatski2020}. The faithful simulation of quantum processing, communication and computation processes is therefore a prerequisite for the viable implementation of such technologies, as it allows for the dynamical testing of hardware specifications and protocol requirements.

To address this demand, a wide variety of classical simulators for quantum networks have been developed. Some target large-scale networks~\cite{NOISE:QuISP2021Flags}, others focus on small-scale high-accuracy simulations~\cite{NOISE:SeQUeNCe2021,NOISE:ReQuSim2024}, and some allow for a hybrid use that addresses both regimes through different simulation modes~\cite{NOISE:NetSquid2021,NOISE:Frank2026}. Additionally, other simulators~\cite{NOISE:SimulaQron2019} support the development of higher-level applications in the upper layers of quantum networks.

However, as these networks scale in size and complexity~\cite{NOISE:Yehia_2024,NOISE:PEREZCASTRO2026112249,NOISE:vanDam_2024,NOISE:Jha2026,NOISE:Brevi2026,NOISE:Wu2024Parallel}, high-accuracy simulations become computationally expensive or strictly intractable with classical computers. To maintain tractability, researchers frequently employ simplification tools to track the evolution of the state. These include, for instance, approximation techniques such as Pauli twirling or randomized compiling~\cite{Noise:PTA_Intro, Wallman2016,Hashim2021}, or restricting the simulation space to stabilizer states and Clifford operations~\cite{NOISE:GKTheroem}.

Although these approximations are computationally efficient and highly effective in specific regimes, particularly for small-scale networks, they can fundamentally limit the ability to capture realistic physical processes. As we transition toward more complex network architectures, these simplifications may fail to account for realistic noise accumulation, potentially leading to qualitatively incorrect predictions of protocol behavior and hardware performance.

\begin{figure}[h]
    \centering
    \includegraphics[width=\linewidth]{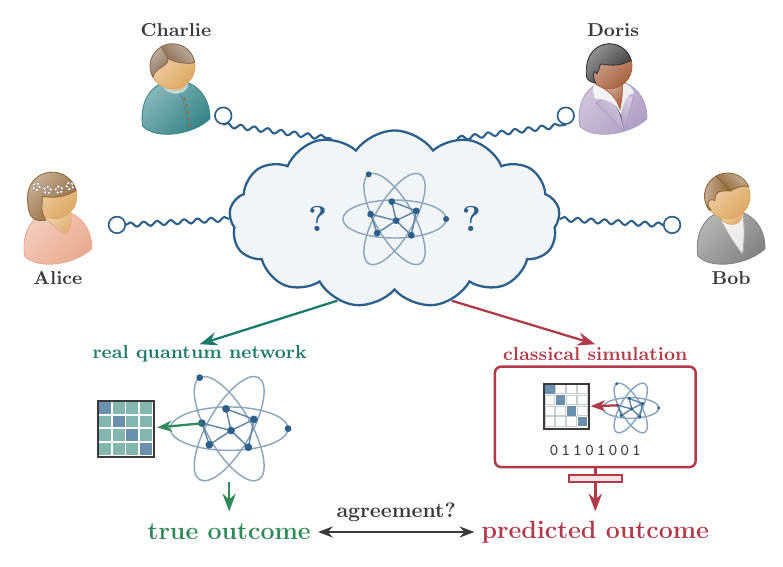}
    \caption{A real quantum network (left) represents the true outcome of a communication task, e.g., successfully establishing a multipartite entangled state between Alice, Bob, Charlie, and Doris, while its classical simulation (right) predicts that outcome using simplified error models. Simplified  assumptions on realistic noise can lead to qualitative and quantitative disagreement between the predicted and true outcomes in either direction.} \label{NOISE:fig:abstract}
\end{figure}

In this work, we demonstrate that such approximations can indeed lead to significant discrepancies in predicting the performance of various quantum communication processes, as illustrated in~\cref{NOISE:fig:abstract}: a real network defines the actual outcome of a task, while the classical simulation based on approximations only predicts it, which may or may not agree with each other. Our results highlight a key finding that typically used simplification techniques can often yield quantitatively and qualitatively inaccurate predictions. We find that these discrepancies can accumulate across multiple protocol steps, leading to significant prediction differences that are fundamentally overlooked by typically used error models.

We analyze these effects in a variety of quantum communication scenarios, ranging from entanglement purification~\cite{NOISE:ReviewEPP2023,Duer2007,DEJMPSPRL1996,Bennett1996, MiguelRamiro2025} and entanglement swapping~\cite{NOISE:swapping} to quantum repeater~\cite{Briegel1998,Dur1999,NOISE:repGen2,NOISE:Luong2016,NOISE:Duan2001,NOISE:Sangouard2011, Meter2013,vanLoock2020, NOISE:reviewRepeater2023} chains in quantum networks. The discrepancies in the predicted behavior can become particularly pronounced in sequential or iterative protocols, which are characteristic of most realistic long-distance quantum communication architectures. Our findings suggest that a rigorous validation of noise models is essential for large-scale simulations of quantum networks, as relying on standard stochastic approximations for decoherence and apparatus imperfections may lead to significantly incorrect predictions of protocol performance and hardware requirements.

This work is structured as follows. We review related works in~\cref{NOISE:sec:previousWork}. We discuss basic noise models and their approximations in~\cref{NOISE:NOISEMODELS}. We show and discuss our analytical and numerical results in~\cref{NOISE:sec:Results}. We finally conclude in~\cref{sec:conclusions}.

\section{Relation to previous work} \label{NOISE:sec:previousWork}
Mapping complex quantum noise channels to simplified stochastic variants, e.g., via the Pauli twirling approximation (PTA) or randomized compiling, is a standard method to ensure classical simulation tractability~\cite{Noise:PTA_Intro, Noise:Alternative_Appr_Method_1}. Alternative approaches to classical simulation of quantum networks include tracking Pauli error propagation~\cite{NOISE:QuISP2021Flags}. 

While these approaches allow for efficient large-scale simulations, it effectively replaces coherent control errors by probabilistic Pauli flips, thereby eliminating coherent contributions (even when accounting for over-rotations) that may accumulate across multiple protocol steps. The question of whether such mappings yield an ``honest'' representation (i.e. a conservative bound on performance thresholds) has been studied in local quantum circuits and error-correcting codes, revealing that PTA can introduce both over-estimations and pessimistic bounds depending on the specific code block and noise distribution~\cite{Noise:Alternative_Appr_Method_1, Noise:Alternative_Appr_Method_2,NOISE:PTA_Alternative_Approch2019, Geller_Nature_Srep, Poulin_PRL_Surface_Code,  Riera-Sabat2025-my, Svore_PRA_Low_Distance,NOISE:Threshold2016,NOISE:ErrorCorr2017,NOISE:PauliQC2026}. Similar conclusions have been obtained in the field of  randomized benchmarking \cite{Proctor17, Epst2014,Wallman2014}.  

Our work shifts this paradigm from static code blocks to dynamic quantum network primitives, differentiating from two recent works. First, Davies \ea~\cite{davies2025accuracytwirledapproximationsrepeater} investigate and derive bounds on the accuracy of the end-to-end fidelity in a linear chain of entanglement swapping events to establish a long-range Bell state for Bell-diagonal and Werner states. Their results show that the approximations based on averaging over all outcomes of Bell-state measurements are either exact for Bell-diagonal states, or highly accurate at high fidelities for Werner states. In contrast, we investigate the concrete limitations of post-selected protocols and the potential propagation of coherent errors through iterative protocols, including entanglement purification. We show that coherent interactions introduce multiple different branches, arising from various outcome sequences, and fidelity oscillations that are completely absent under twirling assumptions. We further observe that averaging over protocol outcomes can significantly deviate from individual outcomes in the presence of off-diagonal noise.

Second, Mondal \ea~\cite{mondal2025efficientformulationquantumnetwork} showed that exact amplitude damping in a linear chain of entanglement swapping events preserves a block-diagonal structure, leading to higher fidelity and average entanglement than Pauli-twirled models. We complement their findings from a simulation perspective: rather than assessing link quality via state twirling, we evaluate the widely used stabilizer tractable \textit{reset channel} gate~\cite{Noise:Alternative_Appr_Method_1,NOISE:resetChannel2,NOISE:resetChannel3} approximation. This approximation is standard in large-scale discrete event quantum network simulators~\cite{NOISE:NetSquid2021, NOISE:SeQUeNCe2021}, which routinely substitute non-Clifford dissipation with Clifford operations or stochastic error-flag tracking to preserve classical simulation tractability.

We find that when amplitude damping occurs during sequential purification and swapping, the reset approximation yields a highly optimistic overestimation of both state fidelity and negativity. Consequently, the sequential composition of local noise approximations does not preserve accuracy in iterative network protocols, making exact validation essential for engineering long-distance quantum internet~\cite{NOISE:Internet_0,NOISE:Internet_1,NOISE:Internet_2,NOISE:Internet_3}.


\section{Noise models}\label{NOISE:NOISEMODELS}
Simulating large-scale quantum networks typically requires simplifying the state space and noise channels to maintain computational tractability. Common techniques include Pauli twirling, which maps arbitrary noise to a diagonal Pauli channel, and the restriction of resources to Bell-diagonal or Werner states. Although these approximations reduce the degrees of freedom in the density matrix, they effectively discard the phase information contained in off-diagonal noise terms. 

We review here the most common noise models considered as well as their approximations, and we analyze them in different quantum communication scenarios. Throughout this manuscript, we define the four Bell basis states $\ket{\Psi_{i,j}}$ as:
\begin{align}
    &\ket{\Psi_{i,j}} = \openone \otimes \mathrm{X}^i \mathrm{Z}^j \ket{\Psi_{0,0}}, \mathrm{~with~} \label{NOISE:BellBasis} \\ \notag
   & \ket{\Psi_{0,0}} = \frac{1}{\sqrt{2}}(\ket{00}+\ket{11}), 
\end{align}
where $\mathrm{Z}$ and $ \mathrm{X}$ are the Pauli gates acting on a single-qubit~\cite{nielsen_chuang_2010}.

\subsection{Over-rotations}
Off-diagonal Pauli terms in channels, which translate to off-diagonal density matrix elements in the Bell basis, are often removed in simplified noise models, for instance through Pauli twirling or diagonal approximations. Although this simplification may be convenient, it neglects the fact that coherent errors may persist and interfere in successive protocol steps. However, such off-diagonal contributions can naturally arise from coherent control errors.

In realistic implementations, such coherent terms can naturally originate from systematic control imperfections, for instance miscalibrated gate parameters or imperfect pulse durations. As a result, their effects can accumulate across multi-step quantum communication protocols, potentially leading to deviations from predictions.

We make use of these over-rotations as a tool to illustrate the general claim of this work, i.e., that simplified error models can lead to wrong predictions. However, this concept is very general and should not be restricted only to the concept of over-rotation introduced here.

We make a distinction between deterministic and probabilistic over-rotations. 

\subsubsection{Deterministic over-rotations}
A common example is a systematic over-rotation with angle $\varphi$ around a Pauli axis $\operatorname{J} \in \{\operatorname{X},\operatorname{Y},\operatorname{Z}\}$, which can be described by the following unitary:
\begin{align}
    \operatorname{U}_{\operatorname{J}}(\varphi) = e^{i \varphi \operatorname{J} } = \cos(\varphi)\openone + \mathrm{i} \sin(\varphi)\operatorname{J}.\label{NOISE:eq:det_overrotation}
\end{align}

As already mentioned, this formulation describes a systematic error that may produce a constant drift in the properties of the resulting state, which can accumulate over successive operations in the network.

\subsubsection{Probabilistic over-rotations}
Alternatively, these over-rotations can also manifest stochastically. Physically, this can, for instance, correspond to a laser source experiencing fluctuations. Mathematically, this is modeled as a completely positive and trace-preserving (CPTP) map:
\begin{align}
    \mathcal{E}_{\operatorname{prob}}(\rho) = (1-p)\rho + p \operatorname{U}_{\operatorname{J}}(\varphi)\rho\operatorname{U}_{\operatorname{J}}^\dagger(\varphi),\label{NOISE:eq:prob_overrotation}
\end{align}
where the over-rotation only occurs with a certain probability $p$. Again, this map gives rise to off-diagonal terms.

\subsection{Noise models approximations}
Over-rotations are an example of coherent errors that generate off-diagonal terms, while many commonly used noise approximations deliberately remove such terms for analytical or numerical convenience. Some examples include:

\subsubsection{Pauli twirling approximation}
The Pauli Twirling Approximation~\cite{Noise:PTA_Intro} is one of the most widely adopted simplification techniques in quantum information processing. Operationally, a general quantum channel $\mathcal{E}$ is mapped to a purely diagonal Pauli channel $\mathcal{E}_{\text{PTA}}$ by averaging the state evolution over the group of $n$-qubit Pauli operators $\mathcal{P}_n$:
\begin{align}
    \mathcal{E}_{\text{PTA}}(\rho) = \frac{1}{4^n} \sum_{\operatorname{P} \in \mathcal{P}_n} \operatorname{P} \mathcal{E}(\operatorname{P} \rho \operatorname{P}) \operatorname{P}.
\end{align}
This operation maps any arbitrary noise process to a convex combination of Pauli errors, effectively eliminating all off-diagonal terms in the Pauli basis, which we refer to as Pauli-diagonal error or noise.

The relevance of the PTA relies on its practical implementability. As a simplification, it maps complex, non-Clifford channels into standard diagonal Pauli channels. This allows one to obtain results via the stabilizer formalism~\cite{NOISE:GKTheroem,Stim2021}, and one can take these types of noise into account by sampling.

As an error mitigation tool for physical hardware, twirling can be actively implemented by interleaving random, local Pauli gates before and after operations, or passively achieved by averaging over specific measurement outcomes in certain tasks~\cite{Wallman2016}.

\subsubsection{Depolarizing noise}
To model noise and imperfections such as memory decoherence, the depolarizing channel is frequently used, as it can be interpreted as a worst case scenario~\cite{Forms2005}. The depolarizing channel probabilistically replaces the state $\rho$ with the maximally mixed state. We use it as an initial noise model for distributed Bell pairs by applying it to one of the qubits (i.e. one of them has been sent through a noisy channel). The depolarizing channel acting on the i-th qubit is defined as: 
\begin{align}
    \mathcal{E}^{(i)}_{\mathrm{dep}}(\rho) = (1-p_{\mathrm{dep}})\,\rho + p_{\mathrm{dep}}\, \mathrm{tr}_i(\rho) \otimes \frac{\openone^{(i)}}{2},\label{NOISE:eq:depol:single}
\end{align}
where $p_{\mathrm{dep}}$ is the error probability. The depolarizing channel is already diagonal in the Pauli basis and contains no off-diagonal elements, in contrast to the over-rotations introduced above. Thus, its fidelity together with the Pauli-diagonal elements remains invariant under the Pauli-twirling approximation.

\subsubsection{Amplitude damping and reset approximations}
Amplitude damping (AD)~\cite{nielsen_chuang_2010} represents a non-Clifford noise channel that models the energy dissipation of a physical system to its ground state. Because it cannot be expressed within the standard stabilizer formalism, an exact classical simulation of large systems subject to AD noise may scale exponentially for some protocols, rendering it practically intractable. The channel is characterized by the following Kraus operator representation:
\begin{align}
    \operatorname{E}_0 &= \ket{0}\bra{0} + \sqrt{1-\gamma} \, \ket{1}\bra{1}, \\
    \operatorname{E}_1 &= \sqrt{\gamma} \, \ket{0}\bra{1},
\end{align}
where $\gamma \in [0,1]$ denotes the decay probability. The application of this channel to an arbitrary single-qubit density matrix yields:
\begin{align}
    \mathcal{E}_{\text{AD}}(\rho) = \begin{pmatrix}
\rho_{00} + \gamma \rho_{11} & \sqrt{1-\gamma} \, \rho_{01} \\
\sqrt{1-\gamma} \, \rho_{10} & (1-\gamma) \rho_{11}
\end{pmatrix}, 
\end{align}
where $\rho_{ij}$ represent the individual entries of the density matrix $\rho$ in the computational basis.

When used as an error model for the initial Bell states we apply the AD channel locally on both qubits.

To overcome the simulation bottlenecks associated with non-Clifford maps, classical quantum network simulators frequently substitute the exact AD channel with tractable approximations~ \cite{NOISE:NetSquid2021, Stim2021, NOISE:SeQUeNCe2021}, such as the reset channel (RC)~\cite{Noise:Alternative_Appr_Method_1,NOISE:resetChannel2,NOISE:resetChannel3}. The reset channel maps a qubit to its ground state $\ket{0}$ with a given probability $\gamma_{\text{RC}}$, and is defined by the following Kraus operators:
\begin{align}
    &\operatorname{M}_0 = \sqrt{1-\gamma_{\text{RC}}} \, \openone, \quad \operatorname{M}_1 = \sqrt{\gamma_{\text{RC}}} \ket{0}\bra{0}, \\
    &\operatorname{M}_2 = \sqrt{\gamma_{\text{RC}}} \ket{0}\bra{1}.
\end{align}
Acting on a general single-qubit density matrix, the reset channel modifies the state according to:
\begin{align}
    \mathcal{E}_{\text{reset}}(\rho) = (1-\gamma_{\text{RC}})\rho + \gamma_{\text{RC}} \ket{0}\bra{0},
\end{align}
which corresponds to a convex combination of the original state and a complete reset to the ground state. There are multiple possible criteria to choose the parameter $\gamma_{\text{RC}}$ in order to approximate an AD-channel with parameter $\gamma$. Here, we pick
\begin{equation}
    \gamma_{\mathrm{RC}} = \frac{1}{3}\left(2 + \gamma - 2\sqrt{1-\gamma}\right),
    \label{NOISE:eq:gammaCJ}
\end{equation}
which makes the Choi--Jamio\l{}kowski fidelity of both channels equal. The Choi--Jamio\l{}kowski fidelity quantifies the ``closeness'' between two quantum processes and entails the most general way to compare both channels. See, e.~g., \cite{Noise:Alternative_Appr_Method_1} for alternative ways to make this choice given additional constraints.

While this mapping is strictly compatible with stabilizer-based simulation frameworks, it fundamentally alters the scaling of the noise approximations. As we demonstrate in subsequent sections, this mathematical simplification can introduce severe discrepancies in the predictions regarding protocols performance and thresholds.

\subsection{Modeling operational noise} 
In this section, we elaborate on our model to describe noisy operations. We treat the initial noise independently of the operation noise, allowing us to attribute the observed noise contribution directly to the appropriate operation. In particular, we model a noisy operation $\Tilde{\mathcal{M}}$ that acts on the state $\rho$ applying the relevant noise channel $\mathcal{E}$ followed by the perfect operation $\mathcal{M}$:
\begin{equation}
\Tilde{\mathcal{M}}(\rho) = \mathcal{M}(\mathcal{E} (\rho)).
\end{equation}

When considering deterministic or probabilistic coherent over-rotation as gate noise, we assume two distinct scenarios. In case of entanglement purification, we model the operational noise as an asymmetric over-rotation of $\varphi$ on Alice's side and $-\varphi$ on Bob's side, which can be interpreted as imperfections or slight calibration offset between their spatially separated setups. For the deterministic over-rotation case this is:
\begin{equation}
    \mathcal{E}_{\text{EPP}}(\rho) = \big(\operatorname{U}_{\text{J}}(\varphi) \otimes \operatorname{U}_{\text{J}}(-\varphi)\big)\, \rho \,\big(\operatorname{U}_{\text{J}}(\varphi) \otimes \operatorname{U}_{\text{J}}(-\varphi)\big)^{\dagger}.\label{NOISE:eq:asymmOffNoise}
\end{equation}
So we model the noisy operations in this case by applying the above channel on both copies, followed by the bilateral CNOT and the measurements. For entanglement swapping operations, all involved qubits are at the same location, so we suppose that the operation causes the same over-rotation of $\varphi$ on both involved qubits. For both scenarios, we compare the operational noise models with their PTA approximations, adjusting the PTA parameters so that the fidelities of both match. 

For the amplitude damping gate noise, we assume that all involved qubits are subject to equal damping with the same parameter $\gamma$, which we compare with the appropriate reset channel, whose parameter $\gamma_{\mathrm{RC}}$ is related to $\gamma$ via~\cref{NOISE:eq:gammaCJ}.

\section{Results}\label{NOISE:sec:Results}
Here, we analyze different quantum communication protocols under the noise models and their approximations considered before. We assume the existence of probabilistic and deterministic over-rotations\footnote{Throughout the entire section, we assume a $\operatorname{Y}$ over-rotation and its PTA approximation acting on the initial Bell state $\ket{\Psi_{0,0}}$, i.e., corresponding to $\operatorname{J}=\operatorname{Y}$ in~\cref{NOISE:eq:prob_overrotation}.}, comparing them with the Pauli-diagonal (twirled) case. We also provide performance comparisons for the AD vs. RC approximation. 

\begin{figure}
    \centering
    \includegraphics[width=\linewidth]{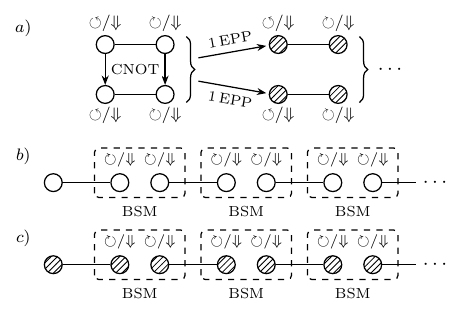}
    \caption{Settings overview: we consider $\ket{\Psi_{0,0}}$ Bell states with initial depolarizing or amplitude damping noise (white circles). Operational noise is either an over-rotation (curved arrows) or amplitude damping noise (double downward arrow). a) Entanglement purification: from two identical initial states, the first pair is probabilistically distilled into a less noisy state (hatched circles) by applying local unitaries, a bilateral \textsc{CNOT} between the two copies (arrow), and local Pauli Z measurements on the second copy; the distillation succeeds, and the first pair is kept, only if the measurement outcomes coincide (00 or 11). For the next EPP round, two identical copies are again assumed, resulting from a successful previous round. b) Entanglement swapping chain: multiple identical initial states are connected via Bell state measurements at intermediate nodes to generate a long-range, but noisy, entangled state. c) Repeater chain: same setting as (b), but with one prior EPP step applied to the initial states.}
    \label{NOISE:fig:setting}
\end{figure}

We present our results for entanglement purification and  entanglement swapping, as well as combination of both within a repeater chain. 

\subsection{Entanglement purification} \label{NOISE:sec:EPP:Results}
We start by analyzing a basic entanglement purification protocol, the \mbox{DEJMPS}~protocol \cite{DEJMPSPRL1996}, see~\cref{NOISE:fig:setting} for our setting. We consider two spatially separated parties, Alice and Bob, who share two identical Bell pairs in each EPP round as input, either noisy initial states or from previous EPP rounds. We assume that equal initial and operational noise affects all or the participating qubits, respectively. For the operation noise, we assume that Alice and Bob experience deterministic over-rotations of~\cref{NOISE:eq:asymmOffNoise} with $\varphi=\pi/64$.

\begin{figure}
    \centering
    \includegraphics[width=0.5\textwidth]{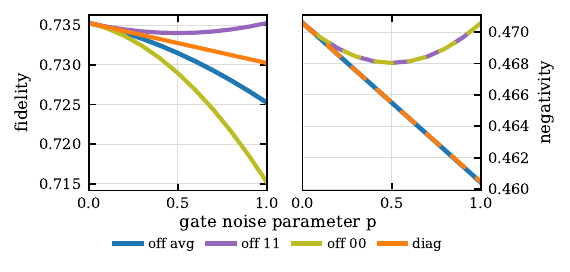}
    \caption{Fidelity (left) and negativity (right) after one \mbox{DEJMPS} step on a $p_{\mathrm{dep}} = 0.4$ depolarized $\ket{\Psi_{0,0}}$ Bell state, with initial fidelity/negativity $0.7$/$0.4$. Operational noise is modeled as a deterministic over-rotation $\mathrm{U_{Y}}(\pm\pi/64)$, with opposite phases on the spatially separated qubits. Curves represent the 00 (olive) and 11 (purple) approximation noise outcomes, their average (blue), and the corresponding PTA (orange).}
    \label{NOISE:plt:DEJMPS:depol:tinyiY:SweepP}
\end{figure}

The results for a single purification step are shown in Fig.~\ref{NOISE:plt:DEJMPS:depol:tinyiY:SweepP}. The first Bell state is kept only upon measurement outcomes 00 or 11, in which case it is left in a less noisy state than before the EPP step. Interestingly, we find that the two outcomes, 00 and 11, differ and lead to two distinct branches (olive and purple lines) as a function of the gate noise parameter $p$. We highlight that this branching behavior of the fidelity is often overlooked, as the ensemble-averaged fidelity is used instead of individual fidelities. Moreover, we observe that the fidelity of the state obtained by averaging the 00 and 11 post-selected outcomes (blue line) differs from the individual results and from the result obtained by the PTA of the operation noise (orange line). These results suggest that the individual branches should be treated separately rather than averaged, especially when the fidelity is the only performance metric. However, the negativity is the same for both averages and for both individual branches. Note the significant differences between curves. We refer to Appendix~\ref{NOISE:intro:EPP} for a detailed analytical analysis of the \mbox{DEJMPS} protocol, where we explain the branching behavior and the discrepancy between the individual branches and their average.

\begin{figure}
    \centering
    \includegraphics{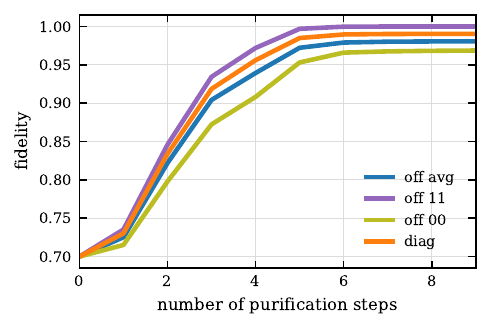}
    \caption{Fidelity over \mbox{DEJMPS} EPP steps for an initial $\ket{\Psi_{0,0}}$ Bell state under depolarizing noise $p_{\mathrm{dep}} = 0.4$. Operational errors are modeled as deterministic $\mathrm{U_{Y}}(\pm \pi/64)$ over-rotations with opposite phases. The 00 (olive) and 11 (purple) approximation noise outcomes converge to slightly different values than their average (blue) and the corresponding PTA (orange). }
    \label{NOISE:plt:DEJMPS:depol:tinyiY:iteration}
\end{figure}

We also analyze the performance of an iterative application of the \mbox{DEJMPS} protocol as a recurrence protocol. Fig.~\ref{NOISE:plt:DEJMPS:depol:tinyiY:iteration} shows the fidelity as a function of the number of purification steps for two trajectories of fixed measurement outcome, 00 (olive) or 11 (purple), their averages (blue) and the corresponding PTA (orange). We yield slightly different asymptotic fidelity results that accumulate in the iterative application of the EPP. Of course, all combinations of measurement outcomes may occur, but we highlight only the most extreme cases.

\begin{figure}
    \centering
    \includegraphics{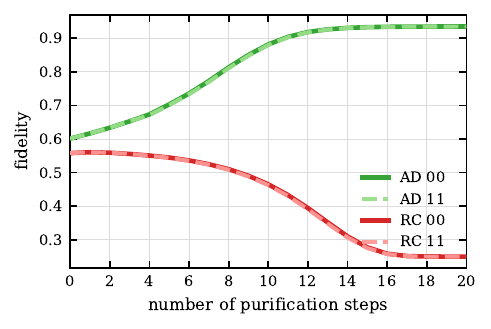}
    \caption{Fidelity over \mbox{DEJMPS} steps for a $\ket{\Psi_{0,0}}$ Bell state subject to initial $\gamma=0.55$ amplitude damping and operational noise modeled as AD with $\gamma_{\mathrm{op}} = 0.05$ (green for 00 and dashed bright green line for 11 outcomes) or the corresponding RC approximation for both initial and operational noise with equivalent parameters (red for 00 and dashed bright red line for 11 outcomes).}
    \label{NOISE:plt:DEJMPS:EPP_AD_RC_CJ}
\end{figure}

Similarly, we analyze how the \mbox{DEJMPS} protocol also shows differences in the amplitude damping vs. reset channel approximation scenario. In~\cref{NOISE:plt:DEJMPS:EPP_AD_RC_CJ}, we plot the fidelity results over the number of \mbox{DEJMPS} steps. We observe that the reset channel approximation predicts the purification protocol to fail, with the fidelity decreasing over successive steps, whereas the exact amplitude damping channel shows the protocol succeeding and converging to high fidelity — the two predictions are thus qualitatively opposite. Across a range of tested initial and gate noise strengths (not shown), we find that the two measurement outcomes 00 and 11 yield the same fidelity and, therefore, do not separate into distinct branches.

\subsection{Entanglement swapping}
For quantum communication, entanglement swapping constitutes another important building block alongside EPP. We investigate how the same noise models and their corresponding approximations perform for this protocol, as detailed in~\cref{NOISE:fig:setting}b. Specifically, we consider a sequence of noisy spatially distributed Bell states, arranged along a linear chain. Except for the endpoints, noisy Bell state measurements at the same spatial position can connect two qubits in independent Bell states, thereby increasing the covering distance.

We first start by investigating a simple scenario to understand the effects of over-rotation in this case. Consider the initial states with $U_\mathrm{Y}(\varphi)$ applied to only the second qubit of each $\ket{\Psi_{0,0}}$ Bell state. We find that, after performing one entanglement swapping operation between two of these states with the result $(i,j)$ (corresponding to projecting on $\ket{\Psi_{i,j}}$) of the Bell state measurement and the corresponding correction back to $\ket{\Psi_{0,0}}$, the resulting state reads as:
\begin{align}
    \rho_{\text{s}}=
    \begin{cases}
    \openone \otimes \operatorname{U}_{\text{Y}}(2\varphi) \ket{\Psi_{0,0}}^{1,4}\bra{\Psi_{0,0}}\openone \otimes \operatorname{U}^{\dagger}_{\mathrm{Y}}(2\varphi) , & \text{if } i=j, \\
    \ket{\Psi_{0,0}}^{1,4}\bra{\Psi_{0,0}}, & \text{if } i \neq j,
\end{cases} \label{NOISE:result:1:swap:iY}
\end{align}
where the different outcomes arise from commuting the correction operations and the noise operators. The resulting state experiences twice the over-rotation noise on the second qubit for $i=j$, or it remains the $\ket{\Psi_{0,0}}$ Bell state for $i\neq j$. We refer to Appendix~\ref{NOISE:app:SWAP:ALL} for details in the derivation.

For multiple steps of entanglement swapping in a chain of Bell states, the over-rotation therefore has the chance to accumulate or cancel out. The probability $p(\theta)$ for ending with $\openone \otimes \operatorname{U}_{\text{Y}}(\theta) \ket{\Psi_{0,0}}$ after $n$ steps of entanglement swapping is given by:
\begin{equation}
    p\left((2k-n+1)\varphi \right) = \frac{1}{2^n} \binom{n}{k}
\end{equation}
for $k\in\{0, 1, \dots, n\}$.

\begin{figure}
    \centering
    \includegraphics{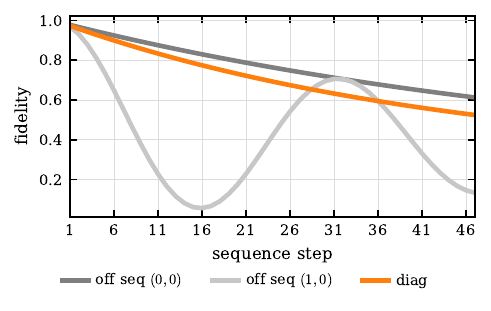}
    \caption{Fidelity over entanglement swapping steps for a $\ket{\Psi_{0,0}}$ Bell state with initial depolarizing noise $p_{\text{dep}}=0.015$. All operations are affected by a $\mathrm{U_{Y}}(\pi/64)$ deterministic over-rotation. Measurement sequences 00 (dark gray) and 10 (light gray) represent the two extreme cases in~\cref{NOISE:result:1:swap:iY}, both deviating from the PTA (orange).}
    \label{fig:swapping}
\end{figure}

For our numerical simulations, we return to our standard error model with initial noise and operational noise. Even in this more involved model, we observe an accumulation of phases. The operational noise model of applying $U_\mathrm{Y}(\varphi)$ to both qubits on which the Bell-state measurement is performed means that at each entanglement swapping step these contributions either accumulate or cancel out. This explains the period of $32$ entanglement swapping steps for $\varphi=\pi/64$ when the measurement outcome is consistently one that accumulates the phases, as shown in Fig.~\ref{fig:swapping}.  So, we observe a general deviation in the predicted fidelity that, for certain measurement outcomes, can accumulate significantly as one adds additional steps to the protocol, e.g., to reach longer distances.  Again, as in the EPP case, we find that fidelity branching occurs for the individual measurement sequences (gray and light gray lines), whereas it does not for the PTA result (orange line). 

\begin{figure}
    \centering
    \includegraphics{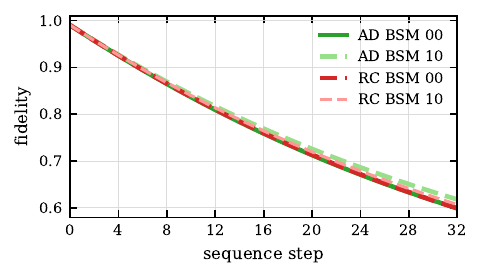}
    \caption{Fidelity over entanglement swapping steps for a $\ket{\Psi_{0,0}}$ Bell state subject to initial $\gamma=0.01$ amplitude damping and operational noise modeled as AD with $\gamma_{\mathrm{op}} = 0.007$ (green) or the corresponding RC approximation for both initial and operational noise with equivalent parameters (red). The Bell state measurement outcomes result into two different outcome sequence branches, \{00,01\} and \{10,11\}; the dark and bright lines show one representative of each, corresponding to outcomes 00 and 10, respectively.
    }
    \label{NOISE:plt:swap:amplitude:gatenoise:fidelity}
\end{figure}

For the amplitude damping and the reset channel approximation, we find that the AD and RC approximation differ only slightly for an iterative entanglement swapping execution, as shown in~\cref{NOISE:plt:swap:amplitude:gatenoise:fidelity}. In addition, we observe that the results are sensitive to the Bell state measurement outcomes. In particular, two distinct branches appear for the sequences of outcomes \{00,01\} and \{10,11\}, we only show one representative for each branch in~\cref{NOISE:plt:swap:amplitude:gatenoise:fidelity} due to legibility reasons.

\subsection{Repeater chain}\label{NOISE:sec:Results:repeaterchain}
\begin{figure}
     \centering
     \includegraphics{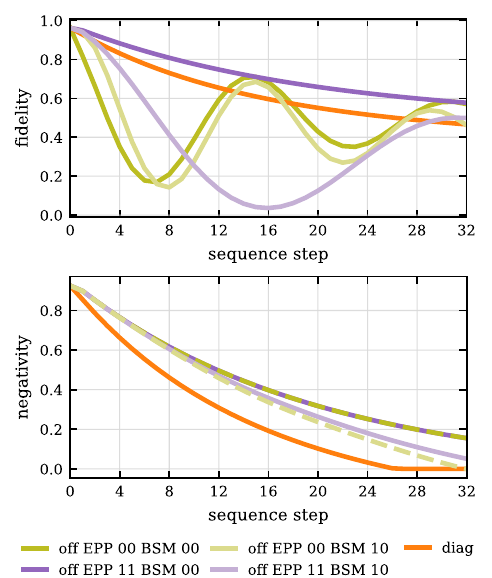}
     \caption{Fidelity (top) and negativity (bottom) over repeater steps for a $\ket{\Psi_{0,0}}$ Bell state subject to an initial depolarizing noise ($p_{\text{dep}}=0.05$). Operational errors are modelled as $\operatorname{U}_{\operatorname{Y}}(\pi/64)$ over-rotation, with opposite phases for the EPP. Colors represent EPP outcomes, 00 (olive) and 11 (purple), and line intensity denotes EPP outcomes, 00 dark and 10 light, and the orange curve shows the PTA. }
     \label{fig:fidelityrepeater}
 \end{figure}
A quantum repeater chain integrates both entanglement purification and entanglement swapping to enable long-range entanglement distribution. We consider the setting illustrated in \cref{NOISE:fig:setting}c, assuming a single entanglement purification step applied to distributed pairs (hatched circles) prior to entanglement swapping, which we denote as a single repeater step. Note that we combine the prior assumptions from the EPP and entanglement swapping.

Our analysis reveals that the repeater chain also exhibits a difference between the outcomes of the \mbox{DEJMPS} protocol and entanglement swapping similar to that observed in the isolated protocols. Crucially, errors originating from both protocols accumulate, resulting in a pronounced divergence in overall performance metrics.

Fig. \ref{fig:fidelityrepeater} shows the state fidelity as a function of the number of repeater steps. We observe how the respective fidelity values for the different cases diverge significantly starting already from the second repeater step. Notably, the fidelity under gate noise with over-rotations exhibits distinct oscillations. To verify that these performance variations do not originate from local unitary rotations, we evaluate the negativity to quantify the distillable entanglement. As shown in \cref{fig:fidelityrepeater}, substantial differences in negativity persist across the different noise models.

\begin{figure}
    \centering
    \includegraphics[width=\linewidth]{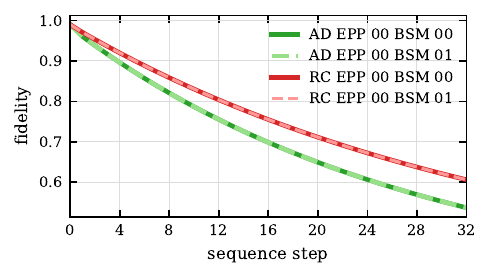}
    \caption{Fidelity over repeater steps for a $\ket{\Psi_{0,0}}$ Bell state subject to $\gamma=0.01$ amplitude damping (green) and its reset channel approximation (red), purified with one \mbox{DEJMPS} step. Operational noise is modeled as AD with $\gamma_{\mathrm{op}} = 0.007$ (green) or its RC approximation (red). The Bell state measurement outcomes result into two different outcome branches, \{00,10\} and \{01,11\}; the dark and bright lines show one representative of each, corresponding to outcomes 00 and 01, respectively. At the small noise magnitudes shown, the branching is not yet resolved.}
    \label{NOISE:fig:repeaterChain:init_AD0.1:AD_p0.02forboth:fid}
\end{figure}

The final scenario under consideration again includes the comparison between amplitude damping and the reset channel, which acts as both initial noise and gate noise. In~\cref{NOISE:fig:repeaterChain:init_AD0.1:AD_p0.02forboth:fid} we plot the state fidelity as a function of the repeater chain steps for the amplitude damping channel alongside its corresponding approximation. We observe that the reset channel systematically overestimates the fidelity of the actual amplitude damping channel. As in the swapping case, the measurement outcomes give rise to distinct branches; however, the prior entanglement purification step changes how the outcomes are grouped: \{00,10\} and \{01,11\}. For the small noise strengths considered here, this branching is not yet resolved, and the curves nearly coincide.

These results highlight our main finding: standard noise approximations may fail to capture non-trivial noise correlations in quantum network protocols, evidencing the necessity of using precise noise models in quantum network simulations. Although we illustrate these effects with simple (artificial) models, they prevail in realistic settings where other noise elements are also involved, as we demonstrate in Appendix~\ref{NOISE:app:realRepeater}.

\section{Conclusion}\label{sec:conclusions}
In this work, we have systematically assessed the limitations of standard noise approximations in the classical simulation of quantum networks affect protocol validation. We demonstrated that coherent errors, such as systematic or probabilistic over-rotations, lead to cumulative interference effects that are fundamentally overlooked when assuming purely Pauli-diagonal models, which e.~g. arise from the Pauli-twirling approximation. In sequential or iterative architectures, such as entanglement purification loops, entanglement swapping chains, and nested quantum repeater networks, these omissions do not merely introduce minor quantitative shifts. Instead, they result in severe qualitative discrepancies, making the result dependent on the measurement outcomes, performance under- and over-estimations, or oscillations in relevant figures of merit. 

Similarly, our comparison between non-Clifford amplitude damping channels and its reset channel approximation reveals that, depending on the protocol, it can also lead to over- and under-estimation of both state fidelity and distillable entanglement in the approximated cases.


These findings underscore a critical caution for the quantum communications community: relying on simplified error models as simulation shortcuts can lead to fundamentally incorrect predictions regarding operational performance. Future work should extend this analysis to further combinations of error models, including asymmetric scenarios, and to more complex networks and tasks, such as more realistic repeater protocols with several purification rounds and multipartite entanglement distribution.


\section*{Acknowledgments}
This research was funded in whole or in part by the Austrian Science Fund (FWF) 10.55776/P36009, 10.55776/P36010, 10.55776/PAT1710825 and 10.55776/COE1. For open access purposes, the author has applied a CC BY public copyright license to any author accepted manuscript version arising from this submission. Finanziert von der Europ\"aischen Union - NextGenerationEU.

\bibliographystyle{apsrev4-2}
\bibliography{main.bib}

\clearpage
\onecolumngrid
\appendix

\section{Entanglement purification}\label{NOISE:intro:EPP}
Entanglement Purification Protocols comprise a set of approaches that allow the distillation of high-fidelity entanglement from multiple noisy states, effectively trading quantity for state quality using only local operations and classical communication (LOCC). The most elementary versions of EPP are recurrence protocols, such as the BBPSSW \cite{Bennett1996} or \mbox{DEJMPS} \cite{DEJMPSPRL1996} protocols, which operate iteratively on a small number of initial pairs to increase fidelity through successive rounds of LOCC. 

In this work, we focus on the \mbox{DEJMPS} protocol, a recurrence-based purification scheme designed to distill high-fidelity entanglement from noisy Bell pairs. Given two identical copies of a state, the protocol proceeds through three stages:

\begin{enumerate} \label{NOISE:DEJMPS:seps}
    \item \textit{Local Rotations.--} Apply the bilateral unitary $\operatorname{U}_1 = \sqrt{-\mathrm{i}\operatorname{X}} \otimes \sqrt{\mathrm{i}\operatorname{X}}$ to both copies.
    \item \textit{Bilateral CNOT.--} Perform a CNOT gate between the two pairs, with the first pair acting as the control and the second as the target.
    \item \textit{Measurement and post-selection.--} Execute local $\operatorname{Z}$-basis measurements on the qubits of the second pair. The first pair is retained only if the results correspond to the $+1$ eigenvalue of $\operatorname{K} = \operatorname{Z} \otimes \operatorname{Z}$ (outcomes $00$ or $11$).
\end{enumerate}

A critical aspect of this assessment is the handling of the measurement outcomes. In experimental reality, the local measurement outcomes are obtained individually and the joint measurement result for $K$ is reconstructed from them. By discarding the individual outcomes one effectively averages over them. We now explicitly distinguish between the $00$ and $11$ outcomes, which is a necessary prerequisite for characterizing the impact of systematic over-rotations. When tracking individual outcomes, off-diagonal noisy terms are not artificially eliminated, and we can identify regimes where these coherent effects lead to significant deviations from the predictions of standard diagonal noise models.

\subsection{Detailed analysis of the \mbox{DEJMPS} protocol} \label{NOISE:App:DEJMPS}
In this Appendix, we investigate how the \mbox{DEJMPS} EPP performs when subject to coherent off-diagonal Pauli. We first elaborate on the difference of the post-measurement state between the individual protocol outcomes and their averages. Subsequently, we analytically analyze the \mbox{DEJMPS} protocol and its performance for coherent off-diagonal Pauli noise in detail.

\subsubsection{Analytical analysis of individual success branches vs averaging}
The successful outcomes of the \mbox{DEJMPS} protocol correspond to the $+1$ eigenvalues of $\operatorname{K}=\operatorname{Z} \otimes \operatorname{Z}$, which are the outcomes that correspond to the projectors $\operatorname{P}_{00}$ and $\operatorname{P}_{11}$ on the $00$ and the $11$ subspaces. We now analyze how averaging these success outcomes differs from the individual ones. We start with expressing the two computational basis states $\ket{00}$ and $\ket{11}$ in the Bell basis:
\begin{align}
    \ket{00} = \frac{1}{\sqrt{2}}(\ket{\Psi_{0,0}}+\ket{\Psi_{0,1}}) \quad \ket{11} = \frac{1}{\sqrt{2}}(\ket{\Psi_{0,0}}-\ket{\Psi_{0,1}}).
    \label{NOISE:eq:CompBasis}
\end{align}
From that follows the post-measurement state in the Bell basis for obtaining the $00$ measurement outcome on the second Bell pair, corresponding to subsystems A2 and B2, as follows:
\begin{align}
    \rho_{00}=&\frac{1}{p_{00}} \mathrm{P}_{00}^{A2,B2} \, \rho \, \mathrm{P}^{\dagger\, A2,B2}_{00} \nonumber \\ 
    =&\frac{1}{2 \, p_{00}} \ket{00}^{A2,B2}\bra{00} \otimes (\bra{\Psi_{0,0}}\rho^{A1,B1}\ket{\Psi_{0,0}}+\bra{\Psi_{0,0}}\rho^{A1,B1}\ket{\Psi_{0,1}} + \bra{\Psi_{0,1}}\rho^{A1,B1}\ket{\Psi_{0,0}}+\bra{\Psi_{0,1}}\rho^{A1,B1}\ket{\Psi_{0,1}}),\nonumber\\
     \label{NOISE:eq:post_state00}
\end{align}
where A1 and B1 denote the subsystems of the first Bell pair. Similarly, for the $11$ measurement result of the second Bell pair, corresponding to subsystems A2 and B2, we obtain the following state in the Bell basis:
\begin{align}
    \rho_{11}=&\frac{1}{p_{11}} \mathrm{P}_{11}^{A2,B2} \, \rho \, \mathrm{P}^{\dagger\, A2,B2}_{11} \\
    =& \frac{1}{2 \, p_{11}} \ket{11}^{A2,B2}\bra{11} \otimes (\bra{\Psi_{0,0}}\rho^{A1,B1}\ket{\Psi_{0,0}}-\bra{\Psi_{0,0}}\rho^{A1,B1}\ket{\Psi_{0,1}} - \bra{\Psi_{0,1}}\rho^{A1,B1}\ket{\Psi_{0,0}}+\bra{\Psi_{0,1}}\rho^{A1,B1}\ket{\Psi_{0,1}}) . \label{NOISE:eq:post_state11}
\end{align}
In both post-measurement states, we see that each contains off-diagonal terms with opposite signs. So, if we average over the both outcomes, we end up with a state that does not contain any off-diagonal density matrix elements:
\begin{align}
    \rho_{\mathrm{avg}} = \frac{p_{00}\operatorname{Tr}_{A2,B2}(\rho_{00})+p_{11}\operatorname{Tr}_{A2,B2}(\rho_{11})}{p_{00}+p_{11}} = \frac{1}{p_{00}+p_{11}} (\bra{\Psi_{0,0}}\rho^{A1,B1}\ket{\Psi_{0,0}}+\bra{\Psi_{0,1}}\rho^{A1,B1}\ket{\Psi_{0,1}}).
\end{align}

\subsubsection{Analytical treatment of a single \mbox{DEJMPS} step}\label{NOISE:App:Calc:DEJMPS}
This appendix analyzes a single EPP step of the \mbox{DEJMPS} protocol, as described in~\cref{NOISE:DEJMPS:seps}, and discusses the corresponding measurement branches. We present this detailed calculation because taking into account the individual phases is crucial for the off-diagonal terms in the Bell state basis, which are of central interest with our noise model. As a first step, we apply the unitary operator $\mathrm{U}_1$ to each Bell state copy, which affects the Bell basis states as follows:
\begin{align}
    \mathrm{U}_1 \ket{\Psi_{a,b}} = (\mathrm{i})^b \ket{\Psi_{a\oplus b, b}},\label{NOISE:eq:DEJMPSU1}
\end{align}
where $\oplus$ corresponds to the addition modulo 2 or the XOR operation. To derive the result of~\cref{NOISE:eq:DEJMPSU1}, we recall that the square root of a Pauli operator is defined by the expression $\sqrt{\pm \mathrm{i} \mathrm{O}} = e^{\pm \mathrm{i} \frac{\pi}{4} \mathrm{O}} = \cos (\frac{\pi}{4}) \mathbf{1} \pm \mathrm{i} \sin (\frac{\pi}{4})\mathrm{O} = \frac{1}{\sqrt{2}}(\mathbf{1} \pm \mathrm{i} \mathrm{O})$. Using this definition, we directly see that $\ket{\Psi_{0,0}}$ remains unchanged under the operator $\operatorname{U}_1$:
\begin{align}
\operatorname{U}_1 \ket{\Psi_{0,0}} &=\sqrt{-\mathrm{i}\operatorname{X}} \otimes \sqrt{\mathrm{i}\operatorname{X}} \,  \ket{\Psi_{0,0}}
= \frac{1}{2} (\openone - \mathrm{i}\operatorname{X}) \otimes  (\openone + \mathrm{i}\operatorname{X}) \ket{\Psi_{0,0}} 
= \frac{1}{2} (\openone \otimes \openone + \mathrm{i}\openone \otimes \operatorname{X} - \mathrm{i} \operatorname{X} \otimes \openone + \operatorname{X} \otimes \operatorname{X}) \ket{\Psi_{0,0}} \nonumber \\
&= \frac{1}{2} \Big( \ket{\Psi_{0,0}} + \mathrm{i} \openone \otimes \operatorname{X} \ket{\Psi_{0,0}} 
- \mathrm{i} \openone \otimes \operatorname{X}^T \ket{\Psi_{0,0}} +  \openone \otimes \operatorname{X}\operatorname{X}^T \ket{\Psi_{0,0}} \Big) 
= \ket{\Psi_{0,0}}, \label{NOISE:eq:app:U1Psi00}
\end{align}
where we used the symmetry of the $\ket{\Psi_{0,0}}$ Bell state to switch the operators from one qubit to the other. For the Bell state $\ket{\Psi_{1,0}}$, the calculation is very similar to~\cref{NOISE:eq:app:U1Psi00}, only an addition $\operatorname{X}$ appears from the definition of $\ket{\Psi_{1,0}}$. The Bell state $\ket{\Psi_{1,1}}$, however, transforms to the state $\mathrm{i} \ket{\Psi_{0,1}}$ under $\operatorname{U}_1$:
\begin{align}
   \operatorname{U}_1 \ket{\Psi_{1,1}} &= \frac{1}{2} \Big( \ket{\Psi_{1,1}} + \mathrm{i} (\openone \otimes \operatorname{XXZ} )\ket{\Psi_{0,0}} -\mathrm{i} (\operatorname{X} \otimes \openone )(\openone \otimes \operatorname{XZ} )\ket{\Psi_{0,0}} + (\operatorname{X} \otimes \openone )(\openone \otimes \operatorname{XXZ} ) \ket{\Psi_{0,0}} \Big) \nonumber \\
   &= \frac{1}{2} (\ket{\Psi_{1,1}} + \mathrm{i}\ket{\Psi_{0,1}} + \mathrm{i}\ket{\Psi_{0,1}} -\ket{\Psi_{1,1}}) = \mathrm{i}\ket{\Psi_{0,1}}, \label{NOISE:eq:app:U1Psi11}
\end{align}
where we used in the last line that we get a minus sign from commuting the $\operatorname{X}$ operator with the $\operatorname{Z}$ operator. Similarly, the Bell state $\ket{\Psi_{0,1}}$ transforms into $\mathrm{i} \ket{\Psi_{1,1}}$, which can be shown analogously to~\cref{NOISE:eq:app:U1Psi11}. Therefore, the operator $\operatorname{U}_1$ transforms an arbitrary density matrix element of a four-qubit quantum state, $\ket{\Psi_{a,b}}_{\mathrm{A1},\mathrm{B1}} \ket{\Psi_{a^{\prime},b^{\prime}}}_{\mathrm{A2},\mathrm{B2}}\bra{\Psi_{c,d}}_{\mathrm{A1},\mathrm{B1}} \bra{\Psi_{c^{\prime},d^{\prime}}}_{\mathrm{A2},\mathrm{B2}}$, as follows:
\begin{align}
    &\operatorname{U}_1 \otimes \operatorname{U}_1 \ket{\Psi_{a,b}}\bra{\Psi_{c,d}} \otimes \ket{\Psi_{a^{\prime},b^{\prime}}}\bra{\Psi_{c^{\prime},d^{\prime}}} \operatorname{U}^{\dagger}_1 \otimes \operatorname{U}^{\dagger}_1 =(\mathrm{i})^{b+b^{\prime}-d-d^{\prime}}\ket{\Psi_{a\oplus b,b}}\ket{\Psi_{a^{\prime}\oplus b^{\prime},b^{\prime}}}\bra{\Psi_{c\oplus d,d}}\bra{\Psi_{c^{\prime}\oplus d^{\prime},d^{\prime}}},\label{eq:u1state}
\end{align}
where we use the result of~\cref{NOISE:eq:DEJMPSU1}. Note that even if we have two identical copies of a state, due to their tensor product, we obtain mixed indices in the density matrix elements. 

The second step in the \mbox{DEJMPS} protocol requires applying a bilateral CNOT gate from the first state copy to the second state copy. Thus, we begin with calculating the result from the first CNOT by expressing the Bell states in the computational basis together with the identity $\operatorname{U}^{\mathrm{A1} \to \mathrm{A2}}_{\mathrm{CNOT}} \ket{m}_{\mathrm{A1}} \ket{n}_{\mathrm{A2}} = \ket{m}_{\mathrm{A1}} \ket{m \oplus n}_{\mathrm{A2}}$:
\begin{align}
    \operatorname{U}&^{\mathrm{B1} \to \mathrm{B2}}_{\mathrm{CNOT}}\ket{\Psi_{i,j}}_{\mathrm{A1},\mathrm{B1}} \ket{\Psi_{k,l}}_{\mathrm{A2},\mathrm{B2}} =\operatorname{U}^{\mathrm{B1} \to \mathrm{B2}}_{\mathrm{CNOT}} \frac{1}{2} (\ket{0}_{\mathrm{A1}} \ket{i}_{\mathrm{B1}} + (-1)^j \ket{1}_{\mathrm{A1}} \ket{1 \oplus i}_{\mathrm{B1}}) \otimes (\ket{0}_{\mathrm{A2}} \ket{k}_{\mathrm{B2}} + (-1)^l \ket{1}_{\mathrm{A2}} \ket{1 \oplus k}_{\mathrm{B2}})  \nonumber \\
    =&\frac{1}{2}\Big(\ket{0}_{\mathrm{A1}} \ket{i}_{\mathrm{B1}}(\ket{0}_{\mathrm{A2}} \ket{k\oplus i}_{\mathrm{B2}} + (-1)^l \ket{1}_{\mathrm{A2}} \ket{1 \oplus k\oplus i}_{\mathrm{B2}}) \nonumber \\
    &+ (-1)^j \ket{1}_{\mathrm{A1}} \ket{1 \oplus i}_{\mathrm{B1}} (\ket{0}_{\mathrm{A2}} \ket{1 \oplus k\oplus i}_{\mathrm{B2}} + (-1)^l \ket{1}_{\mathrm{A2}} \ket{k\oplus i}_{\mathrm{B2}}) \Big).
\end{align}
If we apply the second controlled-NOT to the previous result, we obtain the following result:
\begin{align}
    &\operatorname{U}^{\mathrm{A1} \to \mathrm{A2}}_{\mathrm{CNOT}} \operatorname{U}^{\mathrm{B1} \to \mathrm{B2}}_{\mathrm{CNOT}}\ket{\Psi_{i,j}}_{\mathrm{A1},\mathrm{B1}} \ket{\Psi_{k,l}}_{\mathrm{A2},\mathrm{B2}} = \frac{1}{2} \ket{0}_{\mathrm{A1}} \ket{i}_{\mathrm{B1}} (\ket{0}_{\mathrm{A2}} \ket{k\oplus i}_{\mathrm{B2}} + (-1)^l \ket{1}_{\mathrm{A2}} \ket{1 \oplus k\oplus i}_{\mathrm{B2}}) \nonumber \\
    &+ \frac{1}{2} (-1)^j \ket{1}_{\mathrm{A1}} \ket{1 \oplus i}_{\mathrm{B1}} (\ket{1}_{\mathrm{A2}} \ket{1 \oplus k\oplus i}_{\mathrm{B2}} + (-1)^l \ket{0}_{\mathrm{A2}} \ket{k\oplus i}_{\mathrm{B2}})  = \nonumber \\
    &\frac{1}{2} \Big(\ket{0}_{\mathrm{A1}} \ket{i}_{\mathrm{B1}} (\ket{0}_{\mathrm{A2}} \ket{k\oplus i}_{\mathrm{B2}} + (-1)^l \ket{1}_{\mathrm{A2}} \ket{1 \oplus k\oplus i}_{\mathrm{B2}}) \nonumber \\
    &+ (-1)^{j\oplus l} \ket{1}_{\mathrm{A1}} \ket{1 \oplus i}_{\mathrm{B1}} ((-1)^{l}\ket{1}_{\mathrm{A2}} \ket{1 \oplus k\oplus i}_{\mathrm{B2}} + \ket{0}_{\mathrm{A2}} \ket{k\oplus i}_{\mathrm{B2}})  \Big) = \nonumber \\
    &\ket{\Psi_{i, j \oplus l}}_{\mathrm{A1},\mathrm{B1}} \ket{\Psi_{k \oplus i,l}}_{\mathrm{A2},\mathrm{B2}}, \label{NOISE:eq:BCNOT}
\end{align}
where we used the fact that $(-1)^{-k} = (-1)^{k}$ holds for all integers $k$. With side calculation of~\cref{NOISE:eq:BCNOT}, we obtain for the state after the second step the following:
\begin{align}
   \operatorname{U}^{\mathrm{A1} \to \mathrm{A2}}_{\mathrm{CNOT}} \operatorname{U}^{\mathrm{B1} \to \mathrm{B2}}_{\mathrm{CNOT}} (\operatorname{U}_1 \otimes \operatorname{U}_1) \ket{\Psi_{a,b}}\bra{\Psi_{c,d}} \otimes \ket{\Psi_{a^{\prime},b^{\prime}}}\bra{\Psi_{c^{\prime},d^{\prime}}} (\operatorname{U}^{\dagger}_1 \otimes \operatorname{U}^{\dagger}_1) \operatorname{U}^{\mathrm{B1} \to \mathrm{B2} \dagger}_{\mathrm{CNOT}} \operatorname{U}^{\mathrm{A1} \to \mathrm{A2} \dagger}_{\mathrm{CNOT}}= \nonumber \\
    (\mathrm{i})^{b+b^{\prime}-d-d^{\prime}} \ket{\Psi_{a\oplus b,b\oplus b^{\prime}}}\ket{\Psi_{a\oplus b\oplus a^{\prime}\oplus b^{\prime},b^{\prime}}}\bra{\Psi_{c\oplus d, d\oplus d^{\prime}}}\bra{\Psi_{ c\oplus d \oplus c^{\prime}\oplus d^{\prime},d^{\prime}}}.\label{NOISE:eq:DEJMPSstep2}
\end{align}
As we describe in~\cref{NOISE:DEJMPS:seps}, the final step of the \mbox{DEJMPS} protocol is to measure the second Bell pair, the subsystems $A2$ and $B2$, in the computational basis, regarding $00$ or $11$ as successful outcomes. Thus, we investigate in detail the measurement result $11$ on the subsystem $A2$ and $B2$ of~\cref{NOISE:eq:DEJMPSstep2}, for which we obtain:
\begin{align}
    \bra{11}\ket{\Psi_{a\oplus b\oplus a^{\prime}\oplus b^{\prime},b^{\prime}}} = \frac{1}{\sqrt{2}} (\bra{11}\openone \otimes  \mathrm{X}^{a \oplus b \oplus a^{\prime}\oplus b^{\prime}} \mathrm{Z}^{b^{\prime}} \ket{00}+ \bra{11}\openone \otimes  \mathrm{X}^{a \oplus b \oplus a^{\prime}\oplus b^{\prime}} \mathrm{Z}^{b^{\prime}} \ket{11})  = \frac{1}{\sqrt{2}} \bra{11} \mathrm{Z}^{b^{\prime}} \ket{11} = \frac{1}{\sqrt{2}} (-1)^{b^{\prime}}, \label{NOISE:eq:DEJMPSStep3_11}
\end{align}
where the conditional phase, $(-1)^{b^{\prime}}$, does not appear when measuring $00$. From these considerations in~\cref{NOISE:eq:DEJMPSStep3_11}, we find that the following conditions must be met to have a successful \mbox{DEJMPS} measurement result:
\begin{align}
    a\oplus b\oplus a^{\prime}\oplus b^{\prime} = 0 \nonumber \\
    c\oplus d \oplus c^{\prime}\oplus d^{\prime} = 0, \label{NOISE:eq:constraints:DEJMPS}
\end{align}
which are the same for both outcomes $00$ and $11$, and come from demanding that the $\mathrm{X}$ operator vanishes in~\cref{NOISE:eq:DEJMPSStep3_11}. Therefore, we obtain the following result for the $00$ and $11$ branches after a single \mbox{DEJMPS} protocol step:
\begin{align}
    &\rho^{s=1}_{00} = \frac{1}{2} (\mathrm{i})^{b+b^{\prime}-d-d^{\prime}} \ket{\Psi_{a\oplus b,b\oplus b^{\prime}}}\bra{\Psi_{c\oplus d, d\oplus d^{\prime}}} \nonumber \\
    &\rho^{s=1}_{11} = \frac{1}{2} (\mathrm{i})^{b+b^{\prime}-d-d^{\prime}} (-1)^{b^{\prime} + d^{\prime}} \ket{\Psi_{a\oplus b,b\oplus b^{\prime}}}\bra{\Psi_{c\oplus d, d\oplus d^{\prime}}}. \label{NOISE:eq:DEJMPS:Branches}
\end{align}
With the two conditions of~\cref{NOISE:eq:constraints:DEJMPS} and the expression for the individual branches in~\cref{NOISE:eq:DEJMPS:Branches}, we can identify the elements of the density matrix that pass the selection of the \mbox{DEJMPS} protocol and how they are mapped to other elements.

\subsection{Coherent off-diagonal Pauli noise}\label{NOISE:app:EPP:off-diag}
In this subsection, we investigate the difference between individual successful branches for $\ket{\Psi_{0,0}}$ Bell state subject to coherent off-diagonal Pauli noise on one qubit. In particular, we consider the following single-qubit error channel to describe our error model:
\begin{align}
    \mathcal{D}_{\mathrm{Y}}(\rho) = (1-p) \rho + p \, \operatorname{U}_\mathrm{Y}(\varphi) \rho \operatorname{U}_\mathrm{Y}^\dagger(\varphi) = (1-p) \rho + p (\cos (\varphi) \openone + \mathrm{i} \sin (\varphi)\mathrm{Y})\rho (\cos (\varphi) \openone - \mathrm{i} \sin (\varphi)\mathrm{Y}),
    \label{NOISE:ToyModelY}
\end{align}
where $\varphi$ is the angle that describes the strength of the over-rotation and $p$ is the probability that the error occurs. The deterministic over-rotation scenario from the main text can be recovered by setting $p=1$. The noisy quantum state that we consider is:
\begin{align}
    \rho_{\mathrm{Y}} &= (1-p) \ket{\Psi_{0,0}}\bra{\Psi_{0,0}}+ p \Big( \openone \otimes ( \cos (\varphi)\openone + \mathrm{i}\sin (\varphi)\operatorname{Y})\ket{\Psi_{0,0}} \bra{\Psi_{0,0}} \openone \otimes ( \cos (\varphi)\openone - \mathrm{i} \sin (\varphi)\operatorname{Y}) \Big) \nonumber \\
    &= (1-p + p \cos^2 (\varphi))\ket{\Psi_{0,0}}\bra{\Psi_{0,0}} - p \cos (\varphi)\sin (\varphi) \Big( \ket{\Psi_{0,0}}\bra{\Psi_{1,1}} + \ket{\Psi_{1,1}}\bra{\Psi_{0,0}}\Big) + p \sin^2 (\varphi)\ket{\Psi_{1,1}}\bra{\Psi_{1,1}},\label{NOISE:eq:iY:Overrrotation:DEJMPS}
\end{align}
where we use $\mathrm{i}\operatorname{Y} = \operatorname{ZX}$ to translate the expression in terms of the Bell basis of~\cref{NOISE:BellBasis}. Moreover, we assume that we have two copies of the state given by~\cref{NOISE:eq:iY:Overrrotation:DEJMPS}:
\begin{align}
\rho^{\mathrm{init}}_{\mathrm{Y}} =
&(1-p + p \cos^2 \varphi)^2\,
\ket{\Psi_{0,0}}\bra{\Psi_{0,0}}
\otimes
\ket{\Psi_{0,0}}\bra{\Psi_{0,0}} \nonumber \\
&- (1-p + p \cos^2 \varphi) p \cos \varphi \sin \varphi \,
\Big(\ket{\Psi_{0,0}}\bra{\Psi_{0,0}}
\otimes
\ket{\Psi_{0,0}}\bra{\Psi_{1,1}} + \ket{\Psi_{0,0}}\bra{\Psi_{0,0}} \otimes \ket{\Psi_{1,1}}\bra{\Psi_{0,0}}\Big) \nonumber\\
&+ (1-p + p \cos^2  \varphi) \, p \sin^2 \varphi \,
\ket{\Psi_{0,0}}\bra{\Psi_{0,0}}
\otimes
\ket{\Psi_{1,1}}\bra{\Psi_{1,1}} \nonumber \\
&- (1-p + p \cos^2 \varphi ) p \cos \varphi \sin \varphi \,
\ket{\Psi_{0,0}}\bra{\Psi_{1,1}}
\otimes
\ket{\Psi_{0,0}}\bra{\Psi_{0,0}} \nonumber  \\
&+ p^2 \cos^2 \varphi \sin^2 \varphi \, \Big( \ket{\Psi_{0,0}}\bra{\Psi_{1,1}} \otimes \ket{\Psi_{0,0}}\bra{\Psi_{1,1}} + \ket{\Psi_{0,0}}\bra{\Psi_{1,1}} \otimes \ket{\Psi_{1,1}}\bra{\Psi_{0,0}} \Big) \nonumber  \\
&- p^2 \cos \varphi \sin^3 \varphi \,
\ket{\Psi_{0,0}}\bra{\Psi_{1,1}}
\otimes
\ket{\Psi_{1,1}}\bra{\Psi_{1,1}} \nonumber  \\
&- (1-p + p \cos^2 \varphi ) p \cos \varphi \sin \varphi \,
\ket{\Psi_{1,1}}\bra{\Psi_{0,0}}
\otimes
\ket{\Psi_{0,0}}\bra{\Psi_{0,0}} \nonumber  \\
&+ p^2  \cos^2 \varphi \sin^2 \varphi \, \Big(\ket{\Psi_{1,1}}\bra{\Psi_{0,0}} \otimes \ket{\Psi_{0,0}}\bra{\Psi_{1,1}} +\ket{\Psi_{11}}\bra{\Psi_{0,0}} \otimes \ket{\Psi_{11}}\bra{\Psi_{0,0}}\Big) \nonumber \\
&- p^2 \cos \varphi \sin^3 \varphi \,
\ket{\Psi_{1,1}}\bra{\Psi_{0,0}}
\otimes
\ket{\Psi_{1,1}}\bra{\Psi_{1,1}} \nonumber  \\
&+ (1-p + p \cos^2 \varphi ) p \sin^2 \varphi \,
\ket{\Psi_{1,1}}\bra{\Psi_{1,1}}
\otimes
\ket{\Psi_{0,0}}\bra{\Psi_{0,0}} \nonumber  \\
&- p^2 \cos \varphi \sin^3 \varphi \, \Big( \ket{\Psi_{1,1}}\bra{\Psi_{1,1}} \otimes \ket{\Psi_{0,0}}\bra{\Psi_{1,1}} +\ket{\Psi_{1,1}}\bra{\Psi_{1,1}} \otimes \ket{\Psi_{1,1}}\bra{\Psi_{0,0}} \Big) \nonumber \\
&+ p^2 \sin^4 \varphi \,
\ket{\Psi_{1,1}}\bra{\Psi_{1,1}}
\otimes
\ket{\Psi_{1,1}}\bra{\Psi_{1,1}}.
\end{align}
From the analytical results of the \mbox{DEJMPS} protocol in~\cref{NOISE:eq:constraints:DEJMPS,NOISE:eq:DEJMPS:Branches}, we find that the post-measurement state for the result $00$ is given by:
\begin{align}
    &\rho^{s=1}_{00} \sim  \Big(\frac{1}{2} (1-p + p \cos^2 \varphi )^2 - p^2 \cos^2 \varphi \sin^2 \varphi +\frac{1}{2} p^2 \sin^4 \varphi \Big)\ket{\Psi_{0,0}}\bra{\Psi_{0,0}} \nonumber \\
    &+\mathrm{i} \, p \cos \varphi  \sin \varphi  \Big(1- p +p \cos^2 \varphi - p \sin^2 \varphi \Big) \Big(\ket{\Psi_{0,0}}\bra{\Psi_{0,1}} - \ket{\Psi_{0,1}}\bra{\Psi_{0,0}}\Big) \nonumber \\
    &+ p \sin^2 \varphi \Big( 1- p +2 p \cos^2 \varphi  \Big) \ket{\Psi_{0,1}}\bra{\Psi_{0,1}}, \label{NOISE:DEJMPS:rho00:fidelity:oneStep}
\end{align}
which still contains two non-vanishing off-diagonal terms in the second line. If we consider the post-measurement state corresponding to the result $11$, we find that all off-diagonal terms in the density matrix disappear, leaving us with a linear combination of the following two Bell states:
\begin{align}
        &\rho^{s=1}_{11} \sim \Big(\frac{1}{2} (1-p + p \cos^2 \varphi )^2 + p^2 \cos^2 \varphi \sin^2 \varphi +\frac{1}{2} p^2 \sin^4 \varphi \Big) \ket{\Psi_{0,0}}\bra{\Psi_{0,0}} + (1-p)p \sin^2 \varphi \ket{\Psi_{0,1}}\bra{\Psi_{0,1}}. \label{NOISE:DEJMPS:rho11:fidelity:oneStep}
\end{align}

\subsection{Diagonal Pauli noise}\label{NOISE:app:EPP:diag}
Continuing the previous subsection, we now consider only the diagonal noise contribution
corresponding to the model in~\cref{NOISE:ToyModelY}. This corresponds to the Pauli twirled version of this channel and can be considered as its PTA. The resulting initial state is as follows:
\begin{align}
    \rho_{\mathrm{Y,diag}} = (1-p + p \cos^2 (\varphi))\ket{\Psi_{0,0}}\bra{\Psi_{0,0}} + p \sin^2 (\varphi)\ket{\Psi_{1,1}}\bra{\Psi_{1,1}},\label{NOISE:eq:iY:Overrrotation:DEJMPS:diag}
\end{align}
which yields the following two-copy state:
\begin{align}
\rho^{\mathrm{init}}_{\mathrm{Y,diag}} =
&(1-p + p \cos^2 \varphi)^2\,
\ket{\Psi_{0,0}}\bra{\Psi_{0,0}}
\otimes
\ket{\Psi_{0,0}}\bra{\Psi_{0,0}} \nonumber \\
&+ (1-p + p \cos^2  \varphi) \, p \sin^2 \varphi \,
\left( \ket{\Psi_{0,0}}\bra{\Psi_{0,0}}
\otimes
\ket{\Psi_{1,1}}\bra{\Psi_{1,1}} +\ket{\Psi_{1,1}}\bra{\Psi_{1,1}} \otimes \ket{\Psi_{0,0}}\bra{\Psi_{0,0}}\right) \nonumber \\
&+ p^2 \sin^4 \varphi \,
\ket{\Psi_{1,1}}\bra{\Psi_{1,1}}
\otimes
\ket{\Psi_{1,1}}\bra{\Psi_{1,1}}.
\end{align}
If we use the analytical \mbox{DEJMPS} mapping derived in~\cref{NOISE:eq:DEJMPS:Branches} together with the conditions~\cref{NOISE:eq:constraints:DEJMPS}, we find that the post-measurement state is identical for each measurement outcome and thus
equal to the ensemble-averaged state:
\begin{align}
        &\rho^{s=1}_{\text{diag}} \sim \Big(\frac{1}{2} (1-p + p \cos^2 \varphi )^2 +\frac{1}{2} p^2 \sin^4 \varphi \Big) \ket{\Psi_{0,0}}\bra{\Psi_{0,0}} + (1-p + p \cos^2 \varphi )p \sin^2 \varphi \ket{\Psi_{0,1}}\bra{\Psi_{0,1}}.\label{NOISE:DEJMPS:rhodiag:fidelity:oneStep}
\end{align}

\subsection{Coherent off-diagonal over-rotation: angular dependency}
\begin{figure}
    \centering
    \includegraphics[width=0.7\linewidth]{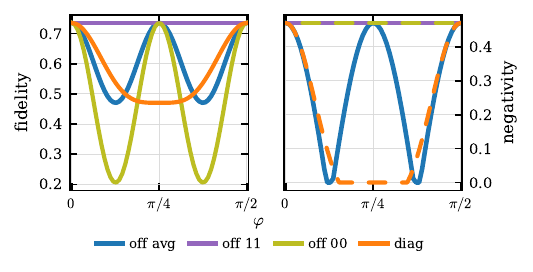}
    \caption{Fidelity (left) and negativity (right) after one \mbox{DEJMPS} step on a $p_{\mathrm{dep}} = 0.4$ depolarized $\ket{\Psi_{0,0}}$ Bell state, with initial fidelity/negativity $0.7$/$0.4$ as a function of the over-rotation angle $\varphi$. Operational noise is modeled as a deterministic over-rotation $\mathrm{U_{Y}}(\varphi)$, with opposite phases on the spatially separated qubits. Curves represent the 00 (olive) and 11 (purple) off-diagonal noise outcomes, their average (blue), and the corresponding PTA (orange).}
    \label{NOISE:plt:DEJMPS:depol:deterministOverort:SweepPhi}
\end{figure}

In~\cref{NOISE:plt:DEJMPS:depol:deterministOverort:SweepPhi}, we present the complementary figure to~\cref{NOISE:plt:DEJMPS:depol:tinyiY:SweepP}. Specifically, we consider the same noisy initial state and vary the over-rotation angle $\varphi$ in the deterministic noise setting ($p=1$). Setting $p=1$ in~\cref{NOISE:DEJMPS:rho11:fidelity:oneStep} eliminates the $\ket{\Psi_{0,1}}\bra{\Psi_{0,1}}$ contribution, leaving the 11 branch independent of the over-rotation angle $\varphi$. For a purly over-rotated initial noise, this results in a constant fidelity of 1 for the 11 outcome. However, in the presence of depolarizing initial noise with $p_{\text{dep}}=0.4$, the branch saturates at a lower angle-independent value, given by $\approx 0.73$. The behavior of the other branch, as well as that of the PTA, can be derived through an analysis analogous to that presented in~\cref{NOISE:app:EPP:off-diag} considering the error model, that we assumed for~\cref{NOISE:plt:DEJMPS:depol:tinyiY:SweepP}.

\section{Entanglement swapping}\label{NOISE:app:SWAP:ALL}
Entanglement swapping is a key operation for extending entanglement over long distances. This is achieved through a Bell State Measurement (BSM) on two qubits, each belonging to an independent entangled pair. It formally corresponds to a projection of a two-qubit system of the orthonormal Bell basis, \cref{NOISE:BellBasis}. Considering two initial Bell states of qubits  (1,2) and (3,4), this projection maps the entanglement from the pairs (1,2) and (3,4) to the pair (1,4). The outcome of the BSM projects the two remaining, distant qubits into one of the four Bell states, followed by a feed-forward unitary correction to reach the target state.

The combination of entanglement swapping and entanglement purification constitutes the operational basis of quantum repeaters. Although swapping extends the entanglement range through concatenated BSMs at intermediate nodes, purification protocols are required to suppress the noise accumulated during transmission and local gate operations. 

\subsection{Analytical treatment of entanglement swapping protocol}\label{app:swapping}
We define the entanglement swapping operator as a projector onto the Bell basis acting on the second and third qubit of a four-qubit state:
\begin{align}
    \operatorname{P}_{i,j} &\coloneqq \openone \otimes \ket{\Psi_{i,j}}\bra{\Psi_{i,j}} \otimes \openone. \label{NOISE:eq:BellPro}
\end{align}

The effect of entanglement swapping, which follows from applying the projector~\cref{NOISE:eq:BellPro} to a tensor product of two $\ket{\Psi_{0,0}}$ Bell states, is given by:
\begin{align}
    &(\openone \otimes \operatorname{P}_{i,j} \otimes \openone ) \ket{\Psi_{0,0}}^{1,2} \ket{\Psi_{0,0}}^{3,4} = (\openone \otimes \ket{\Psi_{i,j}}^{2,3} \otimes \openone ) \bra{\Psi_{0,0}}^{2,3} (\openone \otimes \openone \otimes \openone \otimes \operatorname{X}^i \operatorname{Z}^j) \ket{\Psi_{0,0}}^{1,2} \ket{\Psi_{0,0}}^{3,4} \nonumber \\
    &= (\openone \otimes \ket{\Psi_{i,j}}^{2,3} \otimes \openone ) \frac{1}{\sqrt{2}} \Big( \openone \otimes (\bra{00}^{2,3}+\bra{11}^{2,3}) \otimes \openone \Big) \frac{1}{2} \Big( (\ket{00}^{1,2} +\ket{11}^{1,2}) \otimes  (\openone \otimes \operatorname{X}^i \operatorname{Z}^j \ket{00}^{3,4} + (\openone \otimes \operatorname{X}^i \operatorname{Z}^j \ket{11}^{3,4}) \Big) \nonumber \\
    &= (\openone \otimes \ket{\Psi_{i,j}}^{2,3} \otimes \openone ) \frac{1}{\sqrt{2}} \Big( \openone \otimes (\bra{00}^{2,3}+\bra{11}^{2,3}) \otimes \openone \Big) \frac{1}{2} (\openone \otimes \openone \otimes \openone \otimes \operatorname{X}^i \operatorname{Z}^j) \Big( \ket{0000} + \ket{0011} +  \ket{1100} + \ket{1111} \Big)     \nonumber \\
    &= \frac{1}{2\sqrt{2}}(\openone \otimes \openone \otimes \operatorname{X}^i \operatorname{Z}^j ) \Big( \ket{00}^{1,4} + \ket{11}^{1,4} \Big) \otimes  \ket{\Psi_{i,j}}^{2,3} = \frac{1}{2} \ket{\Psi_{i,j}}^{2,3} \otimes \ket{\Psi_{i,j}}^{1,4} \label{NOISE:app:eq:bellmeasurementoutcome}
\end{align}

\subsection{Coherent off-diagonal Pauli noise}\label{app:swapping:off}
In this subsection, we derive an analytic expression for a single entanglement swapping operation. We consider two identical copies of a Bell state, where the second qubit of each pair is subject to an over-rotation in the Pauli Z direction. Although we carry out the calculation for Pauli Z over-rotations, the corresponding results for the Pauli X and Y over-rotations follow straightforwardly by analogy. We have the following initial state:
\begin{align}
    \rho_{\mathrm{init}} = \operatorname{U}^{\mathrm{T}}_{\mathrm{Z}}(\varphi) \otimes \openone \otimes \openone \otimes \operatorname{U}_{\mathrm{Z}}(\varphi) \ket{\Psi_{0,0}}\bra{\Psi_{0,0}}^{1,2} \ket{\Psi_{0,0}}\bra{\Psi_{0,0}}^{3,4} (\operatorname{U}^{\mathrm{T}}_{\text{Z}}(\varphi) \otimes \openone \otimes \openone \otimes \operatorname{U}_{\mathrm{Z}}(\varphi))^\dagger. \label{NOISE:SWAP:analytic:initial}
\end{align}
If we apply the Bell measurement to the second and third qubit of the state~\cref{NOISE:SWAP:analytic:initial}, we obtain by using~\cref{NOISE:app:eq:bellmeasurementoutcome} the following result:
\begin{align}
    \Big( \openone \otimes \operatorname{P}_{i,j} \otimes \openone \Big) \rho_{\mathrm{init}} \Big( \openone \otimes \operatorname{P}^{\dagger}_{i,j} \otimes \openone \Big) = \Big( \operatorname{U}^{\mathrm{T}}_{\mathrm{Z}}(\varphi) \otimes \operatorname{P}_{i,j} \otimes \operatorname{U}_{\mathrm{Z}}(\varphi) \Big)    \ket{\Psi_{0,0}}\bra{\Psi_{0,0}} \Big( \operatorname{U}^{\mathrm{T}}_{\mathrm{Z}}(\varphi) \otimes \operatorname{P}_{i,j} \otimes \operatorname{U}_{\mathrm{Z}}(\varphi) \Big)^{\dagger} \\
    = \frac{1}{4} \ket{\Psi_{i,j}}^{2,3}\bra{\Psi_{i,j}} \otimes \Big( \operatorname{U}^{\mathrm{T}}_{\mathrm{Z}}(\varphi) \otimes \operatorname{U}_{\mathrm{Z}}(\varphi) \Big) \ket{\Psi_{i,j}}^{1,4}\bra{\Psi_{i,j}}\Big( \operatorname{U}^{\mathrm{T}}_{\mathrm{Z}}(\varphi) \otimes \operatorname{U}_{\mathrm{Z}}(\varphi) \Big)^{\dagger}\\
    = \frac{1}{4} \ket{\Psi_{i,j}}^{2,3}\bra{\Psi_{i,j}} \otimes \Big( \operatorname{U}^{\mathrm{T}}_{\mathrm{Z}}(\varphi) \otimes \operatorname{U}_{\mathrm{Z}}(\varphi) \operatorname{X}^{i}\operatorname{Z}^{j} \Big) \ket{\Psi_{0,0}}^{1,4}\bra{\Psi_{0,0}}\Big( \operatorname{U}^{\mathrm{T}}_{\mathrm{Z}}(\varphi) \otimes \operatorname{Z}^{j} \operatorname{X}^{i} \operatorname{U}_{\mathrm{Z}}(\varphi) \Big)^{\dagger}, \label{NOISE:SWAP:rho2}
\end{align}
where the probability for finding $\ket{\Psi_{i,j}}$ is $p_{i,j}=\frac{1}{4}$. As we want to have the Bell state $\ket{\Psi_{0,0}}$ between qubit 1 and 4, we apply the correction operator $\operatorname{Z}^{j}\operatorname{X}^{i}$ to qubit 4 of the state~\cref{NOISE:SWAP:rho2}. As the $\operatorname{U}_{\mathrm{Z}}(\varphi)$ noise acts on qubit 4, we have to commute the correction operations through, which results in two different cases, depending on the values of $i=0$ and $i=1$:
\begin{align}
    (\operatorname{Z}^{j} \operatorname{X}^{i}) \operatorname{U}_{\mathrm{Z}}(\varphi) \operatorname{X}^{i} \operatorname{Z}^{j} = (\operatorname{Z}^{j} \operatorname{X}^{i}) \Big( \cos(\varphi ) \openone + \mathrm{i} \sin (\varphi )  \operatorname{Z}\Big) \operatorname{X}^{i} \operatorname{Z}^{j} = \begin{cases}
\cos(\varphi ) \openone + \mathrm{i} \sin (\varphi )  \operatorname{Z},  \\
\cos(\varphi ) \openone - \mathrm{i} \sin (\varphi )  \operatorname{Z}, 
\end{cases} =
\begin{cases}
\operatorname{U}_{\mathrm{Z}}(\varphi), & \text{if } i=0, \\
\operatorname{U}_{\mathrm{Z}}(-\varphi), & \text{if } i=1,
\end{cases} \label{NOISE:SWAP:Correction:Z}
\end{align}
where the minus sign in the second case comes from the commutation relation $ZX = -XZ$. This leads to the final state $\rho_3$ on the subsystem of qubit 1 and 4, after renormalization:
\begin{align}
    \rho_{\mathrm{swap,1}}=
    \begin{cases}
    \openone \otimes \operatorname{U}_{\text{Z}}(2\varphi) \ket{\Psi_{0,0}}^{1,4}\bra{\Psi_{0,0}}\openone \otimes \operatorname{U}^{\dagger}_{\mathrm{Z}}(2\varphi) , & \text{if } i=0, \\
    \ket{\Psi_{0,0}}^{1,4}\bra{\Psi_{0,0}}, & \text{if } i=1.
\end{cases} \label{NOISE:SWAP:1stepFinal}
\end{align}
where we shift the noise to qubit four. We conclude from these findings that we obtain a noiseless state in the cases of measuring the states $\ket{\Psi_{1,0}}$ and $\ket{\Psi_{1,1}}$ because the demand for correcting the appearing $X$ operator, which gives us the minus in the commutator and removes the noise. For the case of measuring $\ket{\Psi_{0,0}}$ and $\ket{\Psi_{0,1}}$ the state looks as follows:
\begin{align}
    \rho_{\mathrm{swap,1}, i=0} &= \cos^{2}(2\varphi) \ket{\Psi_{0,0}}^{1,4}\bra{\Psi_{0,0}} + \sin^{2}(2\varphi) \ket{\Psi_{0,1}}^{1,4}\bra{\Psi_{0,1}} \\
    &+\mathrm{i} \sin(2\varphi)\cos(2\varphi) \Big(-\ket{\Psi_{0,0}}^{1,4}\bra{\Psi_{0,1}} + \ket{\Psi_{0,1}}^{1,4}\bra{\Psi_{0,0}}\Big).
\end{align}
Thus, the fidelity with respect to $\ket{\Psi_{0,0}}$ is as follows:
\begin{align}
    F = 
    \begin{cases}
    \cos^{2}(2\varphi) , & \text{if } i=0, \\
    1, & \text{if } i=1.
\end{cases}
\end{align}
If we average across all branches, we obtain an average fidelity of $F_{\text{avg}}=\frac{1}{2}(\cos^{2}(2\varphi)+1)$. Now, we want to derive the results for an X and Y over-rotation. The significant difference occurs in the correction operations of \cref{NOISE:SWAP:Correction:Z}, as the commutation relations change. For a Pauli X over-rotation, the condition in~\cref{NOISE:SWAP:1stepFinal} depends on the value of $j$ instead of $i$, because the $\operatorname{X}$ noise has to commute with the operator $\operatorname{Z}^j$. Hence, we have the following final state:
\begin{align}
    \rho_{\mathrm{swap,1}}=
    \begin{cases}
    \openone \otimes \operatorname{U}_{\text{X}}(2\varphi) \ket{\Psi_{0,0}}^{1,4}\bra{\Psi_{0,0}}\openone \otimes \operatorname{U}^{\dagger}_{\mathrm{X}}(2\varphi) , & \text{if } j=0, \\
    \ket{\Psi_{0,0}}^{1,4}\bra{\Psi_{0,0}}, & \text{if } j=1,
\end{cases} \label{NOISE:SWAP:1stepFinal:X} 
\end{align}
and fidelity with respect to $\ket{\Psi_{0,0}}$ of:
\begin{align}
    F = 
    \begin{cases}
    \cos^{2}(2\varphi) , & \text{if } j=0, \\
    1, & \text{if } j=1.
\end{cases}
\end{align}
For a Pauli Y over-rotation, however, we obtain the conditions that $i=j$ or $i \neq j$ due to the commutations of the Y noise with both correction operators:
\begin{align}
    \rho_{\mathrm{swap,1}}=
    \begin{cases}
    \openone \otimes \operatorname{U}_{\text{Y}}(2\varphi) \ket{\Psi_{0,0}}^{1,4}\bra{\Psi_{0,0}}\openone \otimes \operatorname{U}^{\dagger}_{\mathrm{Y}}(2\varphi) , & \text{if } i=j, \\
    \ket{\Psi_{0,0}}^{1,4}\bra{\Psi_{0,0}}, & \text{if } i \neq j,
\end{cases} \label{NOISE:SWAP:1stepFinal:Y} 
\end{align}
and fidelity with respect to $\ket{\Psi_{0,0}}$ is given by:
\begin{align}
    F = 
    \begin{cases}
    \cos^{2}(2\varphi) , & \text{if } i = j, \\
    1, & \text{if } i \neq j.
\end{cases}
\end{align}
\subsection{Diagonal noise approximation}\label{app:swapping:diag}
For the diagonal noise approximation, we have the following noise channel:
\begin{align}
    D(\rho) = \cos^{2}{\varphi}\rho + \sin^{2}{\varphi} (\openone \otimes \operatorname{Z}) \rho (\openone \otimes \operatorname{Z})^{\dagger}.
\end{align}

If we consider a similar setting as before, two initial Bell states, and shifting the noise such that it acts solely on the first and fourth qubit, we get the following initial state:
\begin{align}
    \rho_{1\text{diag}} &= \cos^{4}{\varphi} \ket{\Psi_{0,0}}\bra{\Psi_{0,0}}^{1,2} \ket{\Psi_{0,0}}\bra{\Psi_{0,0}}^{3,4} + \sin^{4}{\varphi} \ket{\Psi_{0,1}}\bra{\Psi_{0,1}}^{1,2} \ket{\Psi_{0,1}}\bra{\Psi_{0,1}}^{3,4}\nonumber \\
    &+ \cos^{2}{\varphi}\sin^{2}{\varphi} \Big( \ket{\Psi_{0,0}}\bra{\Psi_{0,0}}^{1,2} \ket{\Psi_{0,1}}\bra{\Psi_{0,1}}^{3,4} + \ket{\Psi_{0,1}}\bra{\Psi_{0,1}}^{1,2} \ket{\Psi_{0,0}}\bra{\Psi_{0,0}}^{3,4}\Big).
\end{align}
The state after performing the Bell measurements is given as: 
\begin{align}
    \rho^{\mathrm{swap}}_{1\text{diag}} = (\openone &\otimes \operatorname{P}_{i,j} \otimes \openone) \rho_{1\text{diag}} (\openone \otimes \operatorname{P}^{\dagger}_{i,j} \otimes \openone)= \nonumber \\
    &=\cos^{4}{\varphi} (\openone \otimes \operatorname{P}_{i,j} \otimes \openone) \ket{\Psi_{0,0}}\bra{\Psi_{0,0}}^{1,2} \ket{\Psi_{0,0}}\bra{\Psi_{0,0}}^{3,4} (\openone \otimes \operatorname{P}^{\dagger}_{i,j} \otimes \openone) \nonumber \\
    &+ \sin^{4}{\varphi} (\operatorname{Z}\otimes \operatorname{P}_{i,j} \otimes \operatorname{Z}) \ket{\Psi_{0,0}}\bra{\Psi_{0,0}}^{1,2} \ket{\Psi_{0,0}}\bra{\Psi_{0,0}}^{3,4} (\operatorname{Z}\otimes \operatorname{P}^{\dagger}_{i,j} \otimes \operatorname{Z}) \nonumber \\
    &+ \cos^{2}{\varphi}\sin^{2}{\varphi} (\openone \otimes \operatorname{P}_{i,j} \otimes \operatorname{Z}) \ket{\Psi_{0,0}}\bra{\Psi_{0,0}}^{1,2} \ket{\Psi_{0,0}}\bra{\Psi_{0,0}}^{3,4} (\openone \otimes \operatorname{P}^{\dagger}_{i,j} \otimes \operatorname{Z}) \nonumber \\
    &+ \cos^{2}{\varphi}\sin^{2}{\varphi} (\operatorname{Z}\otimes \operatorname{P}_{i,j} \otimes \openone) \ket{\Psi_{0,0}}\bra{\Psi_{0,0}}^{1,2} \ket{\Psi_{0,0}}\bra{\Psi_{0,0}}^{3,4} (\operatorname{Z}\otimes \operatorname{P}^{\dagger}_{i,j} \otimes \openone) \nonumber \\
    &=\frac{1}{4} \cos^{4}{\varphi} \ket{\Psi_{i,j}}\bra{\Psi_{i,j}}^{1,4} \ket{\Psi_{i,j}}\bra{\Psi_{i,j}}^{2,3} \nonumber \\
    &+\frac{1}{4} \sin^{4}{\varphi} (\operatorname{Z}\otimes \operatorname{Z}) \ket{\Psi_{i,j}}\bra{\Psi_{i,j}}^{1,4} (\operatorname{Z}\otimes \operatorname{Z}) \ket{\Psi_{i,j}}\bra{\Psi_{i,j}}^{2,3} \nonumber \\
    &+\frac{1}{4}\cos^{2}{\varphi}\sin^{2}{\varphi} \Big( (\operatorname{Z} \otimes \openone) \ket{\Psi_{i,j}}\bra{\Psi_{i,j}}^{1,4} (\operatorname{Z} \otimes \openone)+ (\openone \otimes \operatorname{Z}) \ket{\Psi_{i,j}}\bra{\Psi_{i,j}}^{1,4} (\openone \otimes \operatorname{Z}) \Big) \ket{\Psi_{i,j}}\bra{\Psi_{i,j}}^{2,3} \nonumber \\
    &= \frac{1}{4} (\cos^{4}{\varphi}+\sin^{4}{\varphi}) \ket{\Psi_{i,j}}\bra{\Psi_{i,j}}^{1,4} \ket{\Psi_{i,j}}\bra{\Psi_{i,j}}^{2,3} \nonumber \\
    &+\frac{1}{2}\cos^{2}{\varphi}\sin^{2}{\varphi} \Big( (\openone \otimes \operatorname{Z}) \ket{\Psi_{i,j}}\bra{\Psi_{i,j}}^{1,4} (\openone \otimes \operatorname{Z}) \Big) \ket{\Psi_{i,j}}\bra{\Psi_{i,j}}^{2,3},
\end{align}
where the probabilities $p_{i,j}$ to find the Bell state $\ket{\Psi_{i,j}}$ is:
\begin{align}
    p_{i,j} = \operatorname{Tr} \rho^{\mathrm{swap}}_{1\text{diag}} = \frac{1}{4} (\cos^{4}{\varphi}+\sin^{4}{\varphi}) +\frac{1}{2}\cos^{2}{\varphi}\sin^{2}{\varphi} = \frac{1}{4} \Big((\cos^{2}{\varphi}+\sin^{2}{\varphi})^2- 2\cos^{2}{\varphi}\sin^{2}{\varphi}\Big) +\frac{1}{2}\cos^{2}{\varphi}\sin^{2}{\varphi} =\frac{1}{4}.
\end{align}

Analogously as before, when performing the Bell state measurement, we obtain additional operators for which we have to account. Here, we do not get any sign changes from the commutators because we have the same terms on both terms. Thus, the state, at qubit 1 and 4, after correcting the measurement outcome is:
\begin{align}
    \rho^{\mathrm{swap}}_{\mathrm{diag, corr}} &= (\cos^{4}{\varphi} + \sin^{4}{\varphi})\ket{\Psi_{0,0}}\bra{\Psi_{0,0}}^{1,4} +2\cos^{2}{\varphi}\sin^{2}{\varphi} \Big( (\openone \otimes \operatorname{Z}^j \operatorname{X}^i \operatorname{Z}  \operatorname{X}^i \operatorname{Z}^j ) \ket{\Psi_{0,0}}\bra{\Psi_{0,0}}^{1,4} (\openone \otimes \operatorname{X}^i\operatorname{Z}^j \operatorname{Z} \operatorname{Z}^j \operatorname{X}^i) \Big) \nonumber \\
    &= (\cos^{4}{\varphi} + \sin^{4}{\varphi})\ket{\Psi_{0,0}}\bra{\Psi_{0,0}}^{1,4} +2\cos^{2}{\varphi}\sin^{2}{\varphi} \Big( - (\openone \otimes \operatorname{Z}^j  \operatorname{Z}  \operatorname{Z}^j ) \ket{\Psi_{0,0}}\bra{\Psi_{0,0}}^{1,4} (\openone \otimes \operatorname{X}^i \operatorname{Z} \operatorname{X}^i) \Big) \nonumber \\
    &=(\cos^{4}{\varphi} + \sin^{4}{\varphi})\ket{\Psi_{0,0}}\bra{\Psi_{0,0}}^{1,4} +2\cos^{2}{\varphi}\sin^{2}{\varphi} \Big( (\openone \otimes \operatorname{Z}  ) \ket{\Psi_{0,0}}\bra{\Psi_{0,0}}^{1,4} (\openone \otimes \operatorname{Z}) \Big),
\end{align}
and the corresponding fidelity is as follows:
\begin{align}
    F = (\cos^{4}{\varphi} + \sin^{4}{\varphi}),
\end{align}
for all measurement outcomes. We see that the fidelities are different when considering off-diagonal and diagonal noise, respectively. Also, only for the off-diagonal noise case, we find different results for different measurement outcomes. 


\section{Quantum repeaters}\label{NOISE:app:realRepeater}
Quantum repeaters~\cite{Briegel1998,Dur1999} constitute a fundamental architectural primitive in quantum communication, designed to overcome the exponential loss inherent in long-range transmission. By dividing a long-distance channel into shorter, manageable segments, repeaters establish high-fidelity distributed entanglement through the iterative application of entanglement swapping (to extend the range) and entanglement purification (to mitigate decoherence and operational errors). 

\subsection{Realistic repeater chain scenario}
\begin{figure}
    \centering
    \begin{minipage}{0.49\linewidth}
        \centering
        \includegraphics[width=\linewidth]{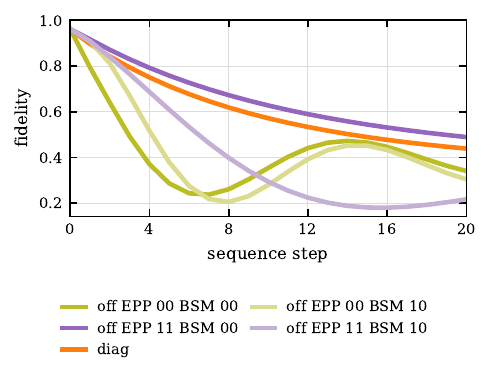}
    \end{minipage}
    \hfill
    \begin{minipage}{0.49\linewidth}
        \centering
        \includegraphics[width=\linewidth]{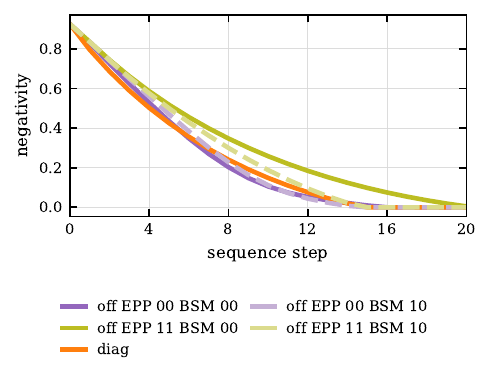}
    \end{minipage}
    \caption{Fidelity (left) and negativity (right) over repeater steps for a $\ket{\Psi_{0,0}}$ Bell state subject to an initial depolarizing noise ($p_{\text{dep}}=0.05$). Operational errors are modeled by a single-qubit depolarizing channel of strength $p_{\text{dep}}=0.003$ followed by an $\operatorname{U}_{\operatorname{Y}}(\pi/64)$ over-rotation, with opposite phases for the EPP. Colors represent EPP outcomes, 00 (olive) and 11 (purple), and line intensity denotes EPP outcomes, 00 dark and 10 light, and the orange curve shows the PTA. }
    \label{NOISE:fig:realisticRepeater}
\end{figure}

Here, we highlight that our findings also hold beyond the assumption of having only a coherent off-diagonal noise for the operations, which we presented in~\cref{NOISE:sec:Results:repeaterchain}. In~\cref{NOISE:fig:realisticRepeater}, we assume the same setting as described in~\cref{fig:fidelityrepeater}, but for the gate noise we apply, in addition to the coherent over-rotation, a single-qubit depolarizing noise with strength $p_{\text{dep}}=0.003$ to each affected qubit.

\end{document}